\documentclass[aip,jcp,11pt]{revtex4-1}
\usepackage[utf8]{inputenc}
\usepackage{color}
\usepackage{amsmath}
\usepackage{amssymb}
\usepackage{graphicx}
\usepackage[makeroom]{cancel}

\usepackage[unicode=true,pdfusetitle,
 bookmarks=true,bookmarksnumbered=false,bookmarksopen=false,
 breaklinks=false,pdfborder={0 0 1},backref=false,colorlinks=false]
 {hyperref}
\usepackage{breakurl}

\usepackage{amsthm}% Useful AMS stuff

\usepackage{tikz}\usetikzlibrary{arrows}
\usepackage{caption}
\usepackage{subcaption}

\usepackage[section]{placeins}

\newcommand{\bra}[1]{\langle #1 \mid}
\newcommand{\ket}[1]{\mid #1 \rangle}
\newcommand{\bracket}[2]{\langle #1 \mid #2 \rangle}

\DeclareMathOperator\erf{erf}
\renewcommand{\imath}{\dot{\iota}}

\newcommand{\DFTBaby}{\textsl{DFTBaby}}

\usepackage{titlesec}\titleformat{\chapter}
{\normalfont\huge\bfseries}{\chaptertitlename\ \thechapter:\ \ }{0em}{} 
\titlespacing*{\chapter}{0pt}{0pt}{30pt}

\begin{document}

\title{DFTBaby: A software package for non-adiabatic molecular dynamics simulations based on long-range corrected tight-binding TD-DFT(B)}

\author{Alexander Humeniuk}
%\author{Jens Petersen}
\author{Roland Mitri\'{c}}
\affiliation{Institut f\"{u}r Physikalische und Theoretische Chemie, Julius-Maximilians Universit\"{a}t W\"{u}rzburg, Emil-Fischer-Stra\ss e 42, 97074 W\"{u}rzburg, Germany}
\email{roland.mitric@uni-wuerzburg.de}

%\abbreviations{}

\begin{abstract}
A software package, called \DFTBaby, is published, which provides the electronic structure needed for running non-adiabatic molecular dynamics simulations at the level of charge-consistent tight-binding DFT. A long-range correction is incorporated to avoid spurious charge transfer states. Excited state energies, their analytic gradients and scalar non-adiabatic couplings are computed using tight-binding TD-DFT. These quantities are fed into a molecular dynamics code, which integrates Newton's equations of motion for the nuclei together with the electronic Schr\"{o}dinger equation. 
Non-adiabatic effects are included by surface hopping. 
As an example, the program is applied to the optimization of excited states and non-adiabatic dynamics of polyfluorene.
The python and Fortran source code is available at \textsl{http://www.dftbaby.chemie.uni-wuerzburg.de}.
\end{abstract}

\maketitle
\begin{enumerate}
\item 
\end{enumerate}

\section{Program Summary}
% Computer Physics Communications                                                                                                                                                                                                             
\begin{itemize}
\item \textsl{Program title:} \DFTBaby
\item \textsl{Licensing provisions:} MIT license
\item \textsl{Programming language:} python and Fortran 90
\item \textsl{Journal Reference:} J. Chem. Phys. 143, 134120 (2015)
\item \textsl{Nature of problem:}
Trajectory-based non-adiabatic molecular dynamics simulations
in excited singlet states for closed-shell molecular systems.
\item \textsl{Solution method:}
The electronic structure is solved using charge-consistent tight-binding DFT
with a long-range correction to avoid spurious charge transfer states.
Excited state energies, their analytic gradients and scalar non-adiabatic couplings
are computed using tight-binding TD-DFT. These quantities are fed in a molecular dynamics code, which integrates Newton's equations of motion for the nuclei together with the electronic Schr\"{o}dinger equation.
Non-adiabatic effects are included by surface hopping.
\end{itemize}

\section{Introduction}
The prediction of the photophysical and photochemical properties of complex materials requires the development of efficient theoretical approaches that allow for the simulation of coupled electron-nuclear dynamics which is induced upon light absorption. In addition, nowadays widely used time-resolved spectroscopy experiments are usually difficult to interpret without resorting to theoretical modeling of the underlying dynamics. 
The versatility of quantum chemistry can be harnessed by using complementary methods: 
Expensive methods can give accurate energetic information for key geometries, while computationally cheaper methods, if applicable, 
can be employed to simulate the non-adiabatic processes directly. 
Molecular dynamics (MD) simulations with trajectories\cite{doltsinis_marx,mitric_2005,mitric_lischka_2006,werner_mitric_2008,barbatti_surface_hopping_review,tavernelli_review,tretiak_semiempirical,prezhdo_dft}, that can hop between different electronic surfaces\cite{tully_surface_hopping}, are particularly popular, 
since each trajectory can be understood as a possible photochemical reaction pathway\cite{lischka_uracil}, while trajectory averages can be directly compared with experimental observables\cite{mitric_werner_2008,betas_furan,polina_harmonic_spectra}. 
% * <roland.mitric@uni-wuerzburg.de> 2017-03-05T10:45:23.644Z:
% 
% Insert here some key references about surface hopping simulations (ours and those from other people like Barbatti&Lischka, Tretiak, Prezhdo etc...)
% 
% ^ <roland.mitric@uni-wuerzburg.de> 2017-03-05T10:46:05.606Z.
In particular, the efficiency of tight-binding DFT\cite{DFTB_elstner,DFTB_for_beginners,TD-DFTB_Niehaus_Elstner_Fraunheim} permits the extension of simulations along different directions: to larger systems,
to longer time-scales or to more trajectories which improves the statistics. 
A vast number of publications exist on surface hopping in combination with all kinds of quantum chemistry methods, among them also DFT\cite{tapavicza_surface_hopping} and tight-binding DFT\cite{dftb_dynamics}.
However, large molecular assemblies pose new problems to DFT(B) and surface hopping, which are absent in smaller molecules: For weakly coupled chromophores, DFT with a local xc-functional predicts unphysically low charge transfer states\cite{dreuw_ZnBc-Bc}. 
Also many degenerate electronic states appear due to excitations localized on the various identical subunits, 
which renders the adiabatic picture partly useless and causes numerical instabilities.
We have recently provided a solution to the problem of erroneous charge transfer by incorporating a long-range correction \cite{lc-tddftb}, an approach that has proven successful in full TD-DFT\cite{long_range_corrected_GGA,long_range_corrected_TDDFT}. 
In order to solve the above mentioned problem connected with the usage of the adiabatic electronic states in the dynamics simulations, the coefficients of the electronic wavefunctions, that determine the surface hopping probabilities, are integrated in a locally diabatic basis, which has been shown to improve numerical stability\cite{local_diabatization_semiempirical,local_diabatization_pyridone}. These two improvements significantly extend the applicability of the TDDFTB to the simulation of ultrafast photodynamics in large systems such as multichromophoric aggregates. 
Arrangements of chromophores that couple only weakly to each other are ubiquitous in nature and technological applications: light-harvesting antennas, dyes and organic photovoltaic devices, to name only a few. If a molecular aggregate contains many identical units, bands of exciton states develop, with many states in a small energy interval. 
As the physical coupling between the states decreases, the energy splitting between the exciton states decreases and the non-adiabatic couplings become more and more peaked. In the extreme case of no diabatic coupling, the exciton states become degenerate and the non-adiabatic couplings turn into $\delta$-functions. Since the labels of adiabatic states are tied to the energetic order, every time two states switch their order, the non-adiabatic coupling exhibits a singularity that ensures the probability for hopping is 100\%. 
This is problematic for numerical integration schemes, as the singularity may be missed if the nuclear time step is not small enough to resolve a peak in the non-adiabatic coupling. Also the integration of the electronic Schr\"{o}dinger equation is unstable if some coupling matrix elements are huge. Granucci et.al. solved this problem by integrating the electronic coefficients in a local diabatic basis
\cite{local_diabatization_semiempirical, local_diabatization_pyridone}.

% Eliminating spurious charge transfer states
%%%%%%%%%%%%%%%%%%%%%%%%%%%%%%%%%%%%%%%%%%%%%%%%%
Dreuw and Head-Gordon analyzed how imprudent application of time-dependent density functional theory (TD-DFT) to weakly interacting molecular systems leads to the prediction of wrong low-lying charge transfer states, that is at odds with electrostatics\cite{dreuw_ZnBc-Bc}: The energy of a charge transfer state should increase as $-R^{-1}$ with the distance between the separated charges. However, the potential energy curve calculated with a pure xc-functional for the charge transfer state of the Zincbacteriochlorin-Bacteriochlorin complex lies below the energy of the lowest valence-excited state and is completely flat as a function of the separation between donor and acceptor moiety. The authors traced this failure of TD-DFT back to the self-interaction error (Coulomb interaction of the electron with its own charge contribution to the total density), which afflicts commonly used density functional \textsl{approximations} but not time-dependent Hartree-Fock (TD-HF) theory.

Maitra \cite{frequency_dependence_ct} showed that the exact exchange-correlation (xc) kernel has a severe frequency dependence. The wrong description of charge transfer states is therefore a consequence of the adiabatic approximation for the xc-functional in TD-DFT. Unfortunately frequency-dependent xc-functionals will probably not be a viable alternative in the near future. The authors of Ref.\cite{long_range_savin} took a more practical approach, noting that DFT is good for exchange- and correlation effects at short distances, while the Hartree-Fock exchange energy has the correct asymptotic limit. They split the Coulomb potential into a short- and long-range part
\begin{equation}
\frac{1}{r} = \underbrace{\frac{1 - \erf\left(\frac{r}{R_{\text{lr}}}\right)}{r}}_{\text{short-range}} + \underbrace{\frac{\erf\left(\frac{r}{R_{\text{lr}}}\right)}{r}}_{\text{long-range}} \label{eqn:split_coulomb}
\end{equation}
where $R_{\text{lr}} \approx 3$ bohr is the distance around which the gradual switch from short- to long-range behaviour happens.
The short-range part of the Coulomb interaction is treated using a local Kohn-Sham density functional, while the long-range part is dealt with wavefunction-based methods, in the simplest case with Hartree-Fock theory.
Iikura \cite{long_range_corrected_GGA} implemented this proposal and Tawada \cite{long_range_corrected_TDDFT} extended it to time-dependent DFT. 

This long-range correction scheme has been recently transferred to tight-binding DFT by Niehaus\cite{lc-dftb_niehaus1, lc-dftb_niehaus2} and to tight-binding TD-DFT by us\cite{lc-tddftb}. Here, we provide the necessary details for the calculation of analytic gradients on excited states (see appendix \ref{sec:analytic_gradients}). The derivations are kept short, because the equations are essentially identical to Refs. \cite{furche, lc_gradients} except for the tight-binding approximations that lead to certain simplifications.

\textbf{Outline of the article:} 
We start with the derivation of the working equations for long-range corrected TD-DFTB and surface hopping with a locally diabatic basis (\ref{sec:theoretical_methods}). These sections are meant largely as a convenient compilation of the known theoretical methods which are implemented in our programs.
The lengthy expressions for analytic gradients and scalar non-adiabatic couplings have been put in the appendix (\ref{sec:analytic_gradients} and \ref{sec:scalar_couplings}). Finally, we demonstrate the scope of application of our programs by computing theoretical absorption and emission spectra for parallel-stacked polyfluorenes with up to five units (section \ref{sec:fluorene}) as well as by simulating the excited state dynamics in these multichromophoric systems.

%%%%%%%%%%%%%%%%%%%%%%%%%%%%%%%%%%%%%%%%%%%%%%%%%%%%%%%%%%%%%%%%%%%%%%%%

\section{Theoretical Methods}
\label{sec:theoretical_methods}
\subsection{Tight-binding TD-DFT}
The working equations of tight-binding DFT are usually derived from a second
order expansion of the DFT energy functional around a reference density that is a superposition of the electron densities of individually neutral atoms\cite{DFTB_elstner,TD-DFTB_Niehaus_Elstner_Fraunheim,DFTB_for_beginners,dftb_review_elstner}.
From an operational point of view, the equations are very similar to semiempirical quantum-chemical methods\cite{semiempirical_methods_A,semiempirical_methods_B,murrell_semiempirical} or charge self-consistent H\"{u}ckel theory\cite{coulson_hueckel_book} with non-orthogonal s-,p- and d-orbitals.
Like in H\"{u}ckel theory, the interaction between atomic orbitals (denoted by Greek letters $\mu,\nu$ etc.) is characterized by a Hamiltonian matrix $H_{\mu\nu}^0$ and the overlap matrix $S_{\mu\nu}$.

The matrix elements depend on geometry and are derived from atomic DFT calculations. 
Matrix elements for atomic valence orbitals of pairs of (pseudo)atoms are calculated in certain orientations ($pp\pi$, $ss\sigma$, $pp\sigma$, etc.) by numerical integration\cite{DFTB_for_beginners}
and are tabulated for all distances. From these tables matrix elements and their gradients can be constructed for all orientations using Slater-Koster rules\cite{slater_koster_rules}.

The electrons occupy molecular orbitals that are linear combinations of the atomic orbitals:
\begin{equation}
 \phi_i(\vec{r}) = \sum_{\mu} C_{\mu i} \phi_{\mu}(\vec{r})
\end{equation}
with the density matrix
\begin{equation}
P_{\mu \nu} = 2 \sum_{i=1}^{N_{\text{elec}}/2} C^*_{\mu i} C_{\nu i}.
\end{equation}

Formation of chemical bonds between different elements causes a redistribution of electronic charge from less to more electronegative atoms. Therefore the total energy contains additional terms for the Coulomb interaction between the partial charges:

\begin{equation}
E_{\text{lc-DFTB}} = \sum_{\mu,\nu} P_{\mu\nu} H_{\mu\nu}^0 + E_{\text{Coulomb}} + E_{\text{exchange}}^{\text{long-range}} + V_{\text{repulsive}}
\end{equation}
The repulsive potential is a sum over atom pairs $(A,B)$ and only depends on the distance $R_{AB}$ between the atoms. It absorbs the interaction between the nuclei and core electrons and is fitted to reproduce DFT energies:
\begin{equation}
V_{\text{repulsive}} = \sum_{A,B} V_{AB}^{\text{rep}}(R_{AB})
\end{equation}
In the strict sense incorporation of the long-range correction would require new fitting of the repulsive potentials. We neglect this and use the same repulsive potentials for calculations with and without the long-range correction.

The residual electron-electron interaction is split into Coulombic interaction at short range and exchange interaction at long range:
\begin{align}
  E_{\text{Coulomb}} &= \frac{1}{2} \sum_{\mu,\sigma,\lambda,\nu} \left(P_{\mu\sigma} - P_{\mu\sigma}^0\right)\left(P_{\lambda\nu}-P_{\lambda\nu}^0\right) (\mu\sigma\vert\lambda\nu) \\
  E_{\text{exchange}}^{\text{long-range}} &= -\frac{1}{4} \sum_{\mu,\sigma,\lambda,\nu} \left(P_{\mu\sigma} - P_{\mu\sigma}^0\right)\left(P_{\lambda\nu}-P_{\lambda\nu}^0\right) (\mu\lambda\vert\sigma\nu)_{\text{lr}}
\end{align}
The 0-th order Hamiltonian $H^0_{\mu\nu}$ already accounts for all interactions between electrons in the neutral atoms. $E_{\text{Coulomb}}$ and $E_{\text{exchange}}$ are the residual Coulomb and exchange energies due to the charge redistribution, which is described by the difference density matrix $\Delta P_{\mu\nu} = P_{\mu\nu} - P^{0}_{\mu\nu}$.
% define reference density matrix
The reference density matrix $P_{\mu \nu}^0$ is a diagonal matrix, since in the reference system
the energy levels $(n,l,m)$ of each atom are occupied as if the atom were isolated:
\begin{equation}
P_{\mu\nu}^0 = \delta_{\mu\nu} \times \text{(occupancy of level $(n_{\mu},l_{\mu},m_{\mu})$ in neutral atom $A_{\mu}$)}
\end{equation}
(Note that in our previous publication\cite{lc-tddftb} in Eqn. (32) we calculated the long-range contribution
using the full density matrix $P_{\mu\nu}$ instead of $\Delta P_{\mu\nu}$, but it turns out that $\Delta P_{\mu\nu}$ is a better choice\cite{kranz_lc-tddftb}.)

Now the tight-binding approximations are made to the 2-electron integrals:
\begin{align}
  (\mu\lambda\vert\sigma\nu) &= \int \int \phi_{\mu}(1) \phi_{\lambda}(1) \frac{1}{r_{12}} \phi_{\sigma}(2) \phi_{\nu}(2) d1 d2 \nonumber \\
                             &\approx \sum_{A,B} \gamma_{AB} q_A^{\mu\lambda} q_B^{\sigma\nu} \label{eqn:tight_binding_approx} \\
  (\mu\lambda\vert\sigma\nu)_{\text{lr}} &= \int \int \phi_{\mu}(1) \phi_{\lambda}(1) \frac{\erf\left(\frac{r_{12}}{R_{\text{lr}}}\right)}{r_{12}} \phi_{\sigma}(2) \phi_{\nu}(2) d1 d2 \nonumber \\
                             &\approx \sum_{A,B} \gamma^{\text{lr}}_{AB} q_A^{\mu\lambda} q_B^{\sigma\nu} \label{eqn:tight_binding_approx_lr}
\end{align}
with the transition charges on atom $A$ (in the atomic orbital basis):
\begin{equation}
q_A^{\mu\lambda} = \frac{1}{2} \left( \delta(\mu \in A) + \delta(\lambda \in A) \right) S_{\mu\nu} \label{eqn:transition_charges_atomic}
\end{equation}
% explain gamma and gamma^lr
The matrices $\gamma_{AB}$ and $\gamma_{AB}^{\text{lr}}$ are defined in Ref.\cite{lc-tddftb}. In short, the $\gamma$-matrices describe the Coulomb interaction between spherically symmetric charge distributions (modelled as Gaussians or Slater functions) centered on the atoms $A$ and $B$. The total charge is smeared out over these charge clouds and amounts to the transition charges assigned to the particular atom according to Eqn. \ref{eqn:transition_charges_atomic}. 3- and 4-center integrals are neglected.
Replacing continuous (transition) densities by atom-centered partial (transition) charges is a very simple form of density fitting\cite{density_fitting} with the spherical Gaussians or Slater functions playing the role of the auxiliary basis functions.

% monopole approximation
This approximation works very well usually, with the exception of $\pi$-electron systems containing heteroatoms. 
For conjugated alternant hydrocarbons the partial charges on the carbons are zero\cite{coulson_hueckel_book} so that
charge self-consistent and non-consistent tight-binding calculations will give the same results.
In the presence of heteroatoms this will not be the case anymore and the assumption that the partial charge cloud
is spherically symmetric becomes a source of error, as evident from the following example:
If a carbon atom is replaced by a heteroatom in an aromatic ring (e.g. turning benzene into pyridine),
the heteroatom will acquire some negative charge.
The charge will be placed in a $\pi$-orbital, that has its maximum above and below the molecular plane and is certainly not spherically symmetric. 

% beyond monopole approximation
%Going beyond the monopole approximation (with spherically symmetric charge clouds) by including partial dipoles or higher multipoles as proposed in \cite{bodrog_thesis} leads to convergence problems in the SCF cycle and no systematic improvement. Similar convergence problems have been reported in Ref.\cite{density_fitting} for small auxiliary basis sets, too. Also, for the minimal basis set of valence orbitals, Mulliken charges can be assigned rather unambiguously to atomic centers. As opposed to this, assigning reasonable partial dipoles without fitting is much more complicated.
% * <roland.mitric@uni-wuerzburg.de> 2017-03-05T11:50:01.937Z:
% 
% We should rethink the part with the convergence problems.
% 
% ^.

% SCF-cycle
Minimizing the total energy $E_{\text{lc-DFTB}}$ under the constraint that the molecular orbitals are orthogonal leads to Kohn-Sham equations for the coefficients $C_{\mu i}$. 
These equations need to be solved self-consistently, since the MO coefficients determine the density matrices, which in turn enter the Coulomb and exchange terms in the energy expression. 

\textbf{Excited states.} 
\label{sec:excitation_energies} 
As in linear-response TD-DFT\cite{casida_equation}, excitation energies $\omega$ of singlet states are obtained from the non-Hermitian eigenvalue problem
\begin{equation}
  \begin{pmatrix}
    \mathbf{A} & \mathbf{B} \\
    \mathbf{B} & \mathbf{A} 
  \end{pmatrix}
  \begin{pmatrix}
    \vec{X} \\ \vec{Y}
  \end{pmatrix}
  = \omega
  \begin{pmatrix}
    \mathbf{1} & 0 \\
    0 & -\mathbf{1} 
  \end{pmatrix}
  \begin{pmatrix}
    \vec{X} \\ \vec{Y}
  \end{pmatrix}
\end{equation}
with 
\begin{align}
  A_{ia,jb} &= \delta_{ij} \delta_{ab} (\epsilon_a - \epsilon_i) + 2(ia|jb) - (ij|ab)_{\text{lr}} \\
  B_{ia,jb} &= 2(ia|jb) - (ib|aj)_{\text{lr}}
\end{align}
after making the tight-binding approximations of Eqns. \ref{eqn:tight_binding_approx} and \ref{eqn:tight_binding_approx_lr} to the 2-electron integrals.

The non-Hermitian eigenvalue problem is solved for the lowest eigenvectors with a Davidson-like iterative algorithm\cite{nonHermitian_Davidson}, which entails the evalulation of matrix products $\mathbf{A}\cdot \vec{v}$ and $\mathbf{B}\cdot \vec{v}$.

The use of a minimal basis set reduces the size of the matrices $\mathbf{A}$ and $\mathbf{B}$, and the use of transition charges speeds up their evaluation. Without long-range correction the evaluation of the matrix products can be performed in a particularly efficient order\cite{heringer}:
\begin{equation}
   (\textbf{A}+\textbf{B})\cdot\vec{v} = 
\left( \begin{array}{c}
  \text{nested sums reduce to matrix} \\
\text{multiplications of lower dimensions}
\end{array} \right)
\end{equation}
With the long-range correction some of the simplicity of the formulae (multiplication of matrices vs. tensor-products) is lost.
With long-range correction the fast execution times needed for MD simulations can still be achieved by restricting the excitations ($i\to a$) to an active space composed of excitations from the highest $M$ occupied to the lowest $N$ virtual molecular orbitals. 
Alternatively the excitation space could be truncated by selecting the single-orbital transitions with the highest oscillator strengths down to a certain threshold as proposed in Ref.\cite{intensity_selection}.
If the active space is chosen reasonably, the only side-effect is a small systematic increase in the excitation energies as shown in the appendix \ref{sec:active_space}.

% Gradients are in appendix \ref{sec:analytic_gradients}
%\subsection{Eliminating spurious charge transfer states}

\begin{figure}[h!]
\includegraphics[width=0.8\textwidth]{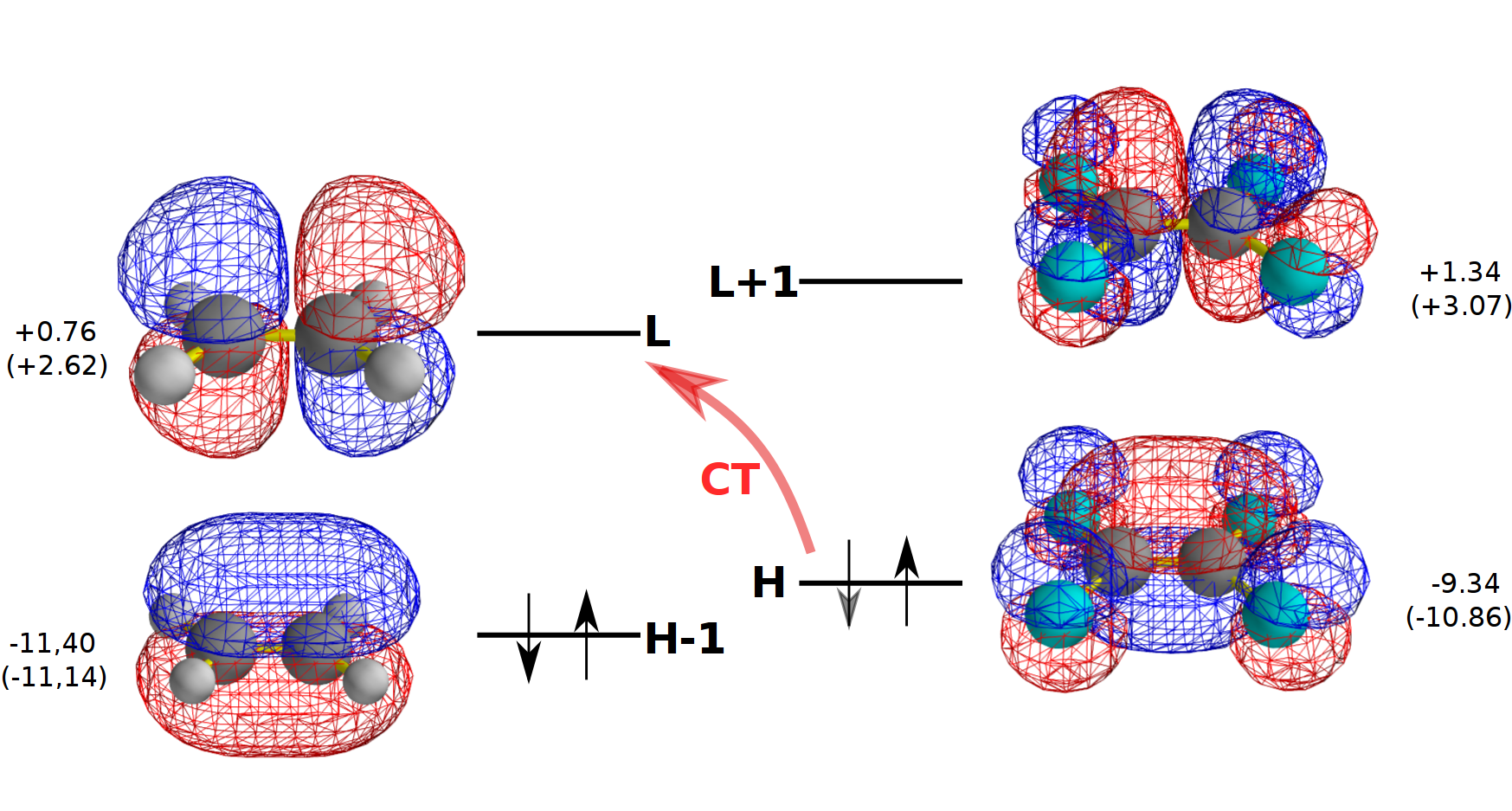}
\caption{Frontier orbitals of ethylene and tetrafluoroethylene. The lc-TD-DFTB orbital energies are given in eV together with the LC-PBE/TZVP energies in brackets for comparison. At large separations the overlap between the HOMO (on tetrafluoroethylene) and the LUMO (on ethylene) vanishes.}
\label{fig:ethylene_tetrafluoroethylene_orbitals}
\end{figure}

In the following, we illustrate the charge-transfer problem in TD-DFT and TD-DFTB and its solution using a long-range correction with an example:
In Ref.\cite{dreuw_ethene_tetrafluoroethylene} Dreuw used a $\pi$-stacked pair of ethylene and tetrafluoroethylene to demonstrate that long-range charge transfer states require non-local exchange. In tetrafluoroethylene the frontier orbitals lie almost 2 eV higher than in ethylene because of the additional nodes between the carbon and fluorine atoms which increase the kinetic energy of the orbitals (see Fig. \ref{fig:ethylene_tetrafluoroethylene_orbitals}).
At small distances the lowest excitation involves charge transfer from the HOMO of tetrafluoroethylene to the LUMO of ethylene. 
As the distance between the molecules increases the energy of the charge transfer should go up to reflect the fact that it costs energy to separate charges, whereas excitations that are localized on either molecule should not depend on the distance. 

Charge transfer over longer distances can be mentally decomposed into three separate steps: ionization of the donor (requiring the ionization energy $IE_{\text{donor}}$), moving the charge to the acceptor molecule against the Coulomb force and adding the electron to the acceptor orbital (releasing the electron affinity $EA_{\text{acceptor}}$). The total energy balance of these steps gives the approximate energy of the charge transfer state:
\begin{equation}
E_{\text{CT}} = IE_{\text{donor}} - EA_{\text{acceptor}} - \frac{1}{R}
\end{equation}

In the TD-DFT(B) picture charge transfer can be viewed as a single excitations from the HOMO (localized on the donor) to the LUMO (localized on the acceptor). If the space of excitation is restricted to only these two orbitals, the long-range TD-DFT excitation energy becomes:
\begin{align}
E_{\text{CT}} = A_{HL,HL} = & \epsilon_L - \epsilon_H + 2(HL|HL) - (HH|LL)_{\text{lr}} \\
           \stackrel{R\to \infty}{\longrightarrow} &   -EA_{\text{acceptor}} -(-IE_{\text{donor}}) + 0 - \frac{1}{R} 
\end{align}
The orbital energies $\epsilon_H$ and $\epsilon_L$ of the HOMO and LUMO, respectively, approximately correspond to minus the ionization ionization energy and electron affinity. The electron-integral $(HL|HL)$ vanishes at large separations, since the HOMO and LUMO are localized on different molecules, so that $\phi_H(r)\phi_L(r) \to 0$. The long-range part of the exchange integral $(HH|LL)_{\text{lr}}$ approaches $\frac{1}{R}$. Without this term, asymptotically the energy of the charge transfer state would be equal to the orbital energy difference.

\begin{figure}[h!]
\includegraphics[width=0.8\textwidth]{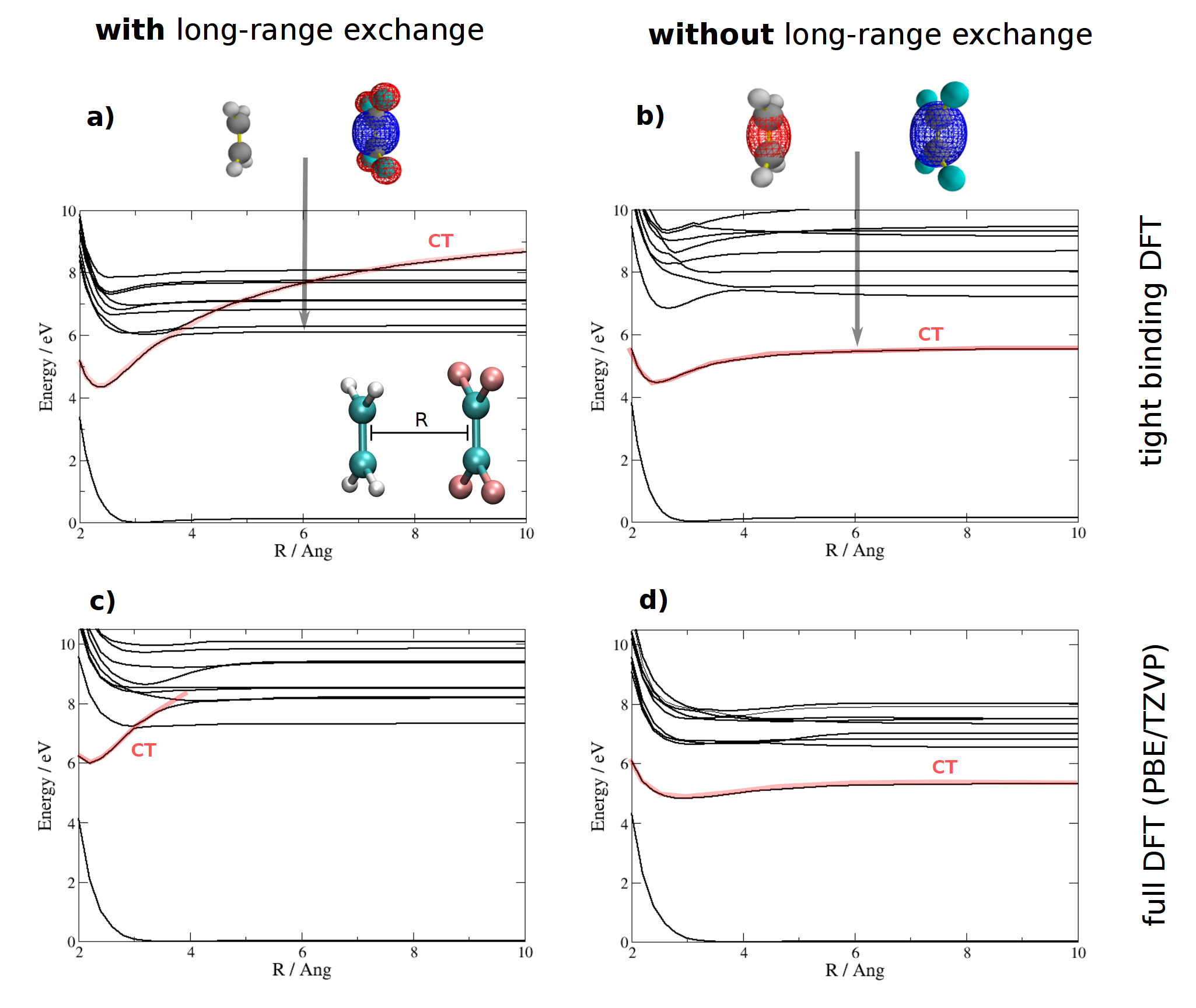}
\caption{Scan of adiabatic potential energy curves for ethylene-tetrafluoroethylene, \textbf{a)} with long-range correction, \textbf{b)} without. 
 \textbf{c)} LC-PBE/TZVP and \textbf{d)} PBE/TZVP. The transition densities in \textbf{a)} and \textbf{b)} show the character of the lowest excited state. Without long-range exchange the lowest state has charge transfer character, whereas it should be a local excitation.}
\label{fig:ethylene_tetrafluoroethylene_comparison}
\end{figure}

In Fig. \ref{fig:ethylene_tetrafluoroethylene_comparison} the potential energies of the lowest 10 excited states are plotted against the distance between the molecular planes.
Since tight-binding DFT is parametrized on the basis of atomic DFT calculations using the PBE functional, the tight-binding results are compared with PBE\cite{density_functional_pbe,g09}/TZVP\cite{basis_set_tzvp} and its long-range corrected version LC\cite{long_range_corrected_GGA}-PBE.
Despite the much lower computational cost, tight binding DFT with and without long-range exchange behaves in the same way as PBE and LC-PBE, respectively:

Without exact exchange the $-\frac{1}{R}$ term is missing, so that the energy of the charge transfer state flattens out as a function of $R$ like the local excitations as soon as the overlap between donor and acceptor molecule goes to zero. In the presence of long-range exchange the charge transfer character has the correct asymptotic $-\frac{1}{R}$ behaviour and cuts through the local excited states whose excitation energy remains constant.

The states with charge transfer character are highlighted in red in Fig. \ref{fig:ethylene_tetrafluoroethylene_comparison} to guide the eye. The difference densities between the 1st excited state and the ground state are shown for a separation of R=$6 $\AA. Without any exact exchange the lowest excitation has charge transfer character for all distances, whereas it should become a local excitation for $R \gg 3$ \AA.

Because of the vanishing overlap between the donor orbital and the acceptor orbital at large separation the transition dipole moment between these two orbitals
vanishes, so that long-range charge transfer states are dark in the absorption spectrum. Although charge transfer states do not show up in the absorption spectrum, they can trap excitations when they are populated indirectly and are very important for organic photovoltaic devices.
The problem of charge transfer states is not limited to the situation where one molecule acts as a donor and the other as an acceptor, so that a charge transfer state is expected in the low energy spectrum.
In fact, any weakly coupled system such as a molecular crystal or a polymer with chromophore units is susceptible to this problem. Without long-range exchange the spectrum will be contaminated by unphysical charge transfer states
that lie below the lowest local excitation. 
%If two arbitrary identical molecules are placed infinitely far apart, so that they do not interact in any way, the local HOMO-LUMO excitation will be degenerate with a charge transfer from the HOMO of one molecule to the LUMO of the other. 
Consider two arbitrary identical molecules that are placed infinitely far apart, so that they do not interact in any way. The frontier orbitals of	the combined system are	linear combinations of the individual HOMO and LUMO orbitals on	each molecule and are delocalized over both molecules. They can be localized on each molecule, leading to 4 frontier orbitals: HOMO(1) and LUMO(1) localized on the first molecule and HOMO(2) and LUMO(2) localized on the second one. The local HOMO(1) $\to$ LUMO(1) excitation will be degenerate with a charge transfer from the HOMO(1) of the first molecule to the LUMO(2) of the other.
This jump of a charge over a very long distance is an obviously unphysical artifact of density functional approximations that neglect exact long-range exchange. 

These problems will also appear in the simulations of excited state nonadiabatic dynamics where the unphysical low-lying charge transfer states will lead to artificial nonradiative-relaxation channels. Therefore, the simulations of nonadiabatic dynamics in molecular aggregates at the TDDFT(B) level should be generally performed only in combination with long-range correction.
\FloatBarrier

\subsection{Surface hopping}
\label{sec:surface_hopping}
Tully's surface hopping\cite{tully_surface_hopping} is a stochastic method for simulating non-adiabatic events in molecular dynamics that takes place on multiple electronic potential energy surfaces. Despite its successes it cannot be derived rigorously from the time-dependent variational principle or any other principle. It is an ad hoc procedure that works very well in practice and combines well with quantum chemistry methods, that make the Born-Oppenheimer separation between fast electronic (denoted by $\vec{r})$ and the slow nuclear degrees of freedom (denoted by $\vec{R}$). 

For a fixed nuclear geometry $\vec{R}$, a quantum chemistry code gives a manifold of electronic wavefunctions $\Psi_i(\vec{r};\vec{R})$  with adiabatic energies $\{E_i(\vec{R})\}_{i=1,2,\ldots}$, that depend parametrically on the nuclear coordinates. The nuclear wavefunction is kept out of the equation but its hidden presence manifests itself in the form of Berry phases: When the electronic wavefunction is transported adiabatically around a point of energetic degeneracy (conical intersection) back to its starting location, it acquires a sign change, which would be cancelled by the phase of the nuclear part of the wavefunction.

In surface hopping, the nuclear wavefunction is approximated by a delta-function, or point $(\vec{R},\vec{P})$ in phase space. An ensemble of trajectories drawn from some distribution $f(\vec{R},\vec{P})$ (Wigner distribution, Boltzmann distribution, etc.) can be given different interpretations: either as the finite spread of the quantum-mechanical wavepacket due to the Heisenberg uncertainty principle and/or the uncertainty about the phase space positions of the classical nuclei due to the finite temperature\cite{barbatti_initial_condition_sampling}. This ambiguity makes surface hopping a perfect match for simulations at room temperature, which usually are affected by both types of uncertainty.

The electrons exert forces on the nuclei, which are different for each electronic Born-Oppenheimer state $i$. Assuming the electrons are in state $c$, the equation of motion for the classical nucleus $A$ is given by Newton's equation:
\begin{equation}
m_A \ddot{\vec{R}}_A = - \vec{\nabla}_A E_c \label{eqn:newton_equation}
\end{equation}
This defines a nuclear trajectory $\vec{R}(t)$ propagating on the ''current'' electronic state $c$. When electronic states come close in energy or cross, the Born-Oppenheimer separation breaks down and transitions between electronic states need to be considered. A trajectory is restricted to move on one surface at a time, but a sudden hop can transfer it to another surface, leading to a discontinuity in the acceleration.
To determine the propensity to switch to another state, the nuclear trajectory is equipped with an electronic wavefunction, that is a linear combination of the instantaneous adiabatic eigenstates:
\begin{equation}
\Psi(\vec{r};\vec{R}(t)) = \sum_k C_k(t) \Psi_k(\vec{r};\vec{R}(t)) \label{eqn:electronic_ansatz}
\end{equation}
The modulus squared of a coefficient, $\vert C_k(t) \vert^2$, gives the probability of the trajectory to move on the potential energy surface $k$. When a trajectory starts initially on the electronic state $i$, then $C_k(t=0) = \delta_{ki}$. The time-evolution of the coefficients along the trajectory is governed by a Schr\"{o}dinger-like differential equation. To make clear which non-adiabatic effects are included and which are not, it will be derived in some length.

The Hamiltonian of the total system 
\begin{equation}
\hat{H} = \hat{T}_{\text{nuc}} + \hat{H}_{\text{elec}}(\vec{r};\vec{R})
\end{equation}
is split into the nuclear kinetic energy $\hat{T}_{\text{nuc}} = \sum_{A=1}^{N_{\text{at}}} - \frac{\hbar^2}{2 m_A} \vec{\nabla}_A^2$ and the electronic Hamiltonian $\hat{H}_{\text{elec}}(\vec{r};\vec{R})$ that comprises the remaining interactions that do not depend on the nuclear momenta. The electronic wavefunctions $\Psi_k(\vec{r};\vec{R}(t))$ are the eigenfunctions of the electronic Hamiltonian,
\begin{equation}
\hat{H}_{\text{elec}} \Psi_k(\vec{r};\vec{R}(t)) = E_k \Psi_k(\vec{r};\vec{R}(t)),
\end{equation}
but not of the total Hamiltonian, since
\begin{equation}
\bra{\Psi_i} \hat{H} \ket{\Psi_j} = \sum_{A=1}^{N_{\text{at}}} -\frac{\hbar^2}{2 m_A} \bra{\Psi_i}\vec{\nabla}_A^2 \ket{\Psi_j} + \delta_{ij} E_j(\vec{R}).
\end{equation}
The first term involving $\vec{\nabla}^2$ (the diagonal Born-Oppenheimer correction) is neglected, not necessarily because it is small, but simply because this quantity is not readily available from quantum chemistry codes. Its inclusion in surface hopping methods has actually been shown to lead to inferior results for strongly coupled potential energy surfaces\cite{izmaylov_dboc}.

When subsituting the ansatz in Eqn. \ref{eqn:electronic_ansatz} into the electronic time-dependent Schr\"{o}dinger equation $\imath \hbar \frac{\partial}{\partial t} \ket{\Psi} = \hat{H}_{elec} \ket{\Psi}$ one needs to keep in mind the parametric dependence of $\Psi_i$ on $\vec{R}(t)$:
\begin{equation}
\imath \hbar \sum_{i=1} \left( \frac{dC_i}{dt} \ket{\Psi_i} + C_i \ket{\vec{\nabla}_{\vec{R}} \Psi_i} \cdot \frac{d \vec{R}}{dt} \right) = \sum_i C_i(t) E_i(\vec{R}(t)) \ket{\Psi_i}
\end{equation}
Multiplication from the left with $\bra{\Psi_j(\vec{R}(t))}$ and using the orthogonality of electronic states at the same nuclear geometry, $\bracket{\Psi_j(\vec{R}(t))}{\Psi_i(\vec{R}(t))} = \delta_{ji}$, gives:
\begin{equation}
\imath \hbar \frac{d C_j}{dt} = \sum_i \left(E_i(\vec{R}(t)) \delta_{ji} - \imath \hbar \bracket{\Psi_j}{\vec{\nabla}_{\vec{R}} \Psi_i} \cdot \frac{d \vec{R}}{dt} \right) C_i(t) \label{eqn:electronic_se}
\end{equation}
Although the nonadiabatic coupling vector $\bracket{\Psi_j}{\vec{\nabla}_{\vec{R}} \Psi_i}$ appears in the Eqn. \ref{eqn:electronic_se} its calculation is not needed for the propagation of the electronic degrees of freedom since the scalar product between the non-adiabatic coupling vector and the nuclear velocity vector $\frac{d \vec{R}}{dt}$ can be approximated by overlaps between electronic wavefunctions at successive nuclear time steps\cite{tavernelli_tddft_wavefunction}:
\begin{equation}
\begin{split}
\bracket{\Psi_j}{\vec{\nabla}_{\vec{R}} \Psi_i} \cdot \frac{d\vec{R}}{dt} &= \bracket{\Psi_j}{\frac{\partial}{\partial t} \Psi_i} \\
  &\approx \frac{1}{2 \Delta t} \left(\bracket{\Psi_j(\vec{r};\vec{R}(t))}{\Psi_i(\vec{r};\vec{R}(t+\Delta t))} - \bracket{\Psi_j(\vec{r};\vec{R}(t+\Delta t))}{\Psi_i(\vec{r};\vec{R}(t))} \right)
\end{split}
\end{equation}
Expressions for calculating these scalar couplings between singlet TD-DFTB ``wavefunctions`` are listed in appendix \ref{sec:scalar_couplings}.

The integration of Newton's equation \ref{eqn:newton_equation} (with a time step of $\Delta t \approx$ 0.1 fs ) and the electronic Schr\"{o}dinger equation \ref{eqn:electronic_se} (with a much smaller time step $\Delta t_{\text{elec}} \approx 10^{-5}$ fs) are intertwined. After each nuclear time step, the electronic density matrix,
\begin{equation}
\rho_{kl}(t) = C_k^*(t) C_l(t)
\end{equation}
is calculated and the probability for changing the current electronic state from $i$ to $j$ is 
calculated according to the formula\cite{hopping_probabilities}:
\begin{equation}
P_{i\to j}=\Theta(-\dot{\rho}_{ii})\Theta(\dot{\rho}_{jj})\frac{\left(-\dot{\rho}_{ii}\right)\dot{\rho}_{jj}}{\rho_{ii}\sum_{k}\Theta(\dot{\rho}_{kk})\dot{\rho}_{kk}}\Delta t
\end{equation}
where $\Theta(x)$ is the Heaviside step function, that is $1$ for $x \ge 0$ and $0$ otherwise.
This formula is an improvement over Tully's original fewest switches formula, since it also considers rates of change. The diagonal elements of the density matrix, $\rho_{kk}(t)$, are called the \textsl{quantum populations}. The unprovable tenet of surface hopping is that the average numbers of trajectories on each 
electronic state (the \textsl{trajectory populations}) approach the quantum populations in the limit of a very large ensemble of trajectories.

The off-diagonal elements $\rho_{kl}$ are called \textsl{quantum coherences}. The lack of a nuclear wavefunction leads to the phenomenon of over-coherence: After a trajectory leaves a region of strong non-adiabatic coupling, the induced coherences do not decay but remain constant. 

\textbf{Momentum Rescaling.} During a surface hop the potential energy has a discontinuity, unless a surface hop occurs at a conical intersection. To restore energy conservation 
the momentum is rescaled uniformly $(\vec{p} \to s \vec{p})$ so that the change in kinetic energy $T$ offsets the change in potential energy caused by the hop from $i$ to $j$:
\begin{equation}
E_i + T = E_j + s^2 T.
\end{equation}
If the quadratic equation for the scaling factor $s$,
\begin{equation}
s = \sqrt{1 + \frac{E_i - E_j}{T}},
\end{equation}
does not have a real solution, the surface hop is rejected and the trajectory continues on the old potential energy surface. This happens when a slow trajectory attempts to hop to a higher energy level that could not be reached even if all kinetic energy would be converted to potential energy.

% * <roland.mitric@uni-wuerzburg.de> 2017-03-07T10:59:17.846Z:
% 
% Describe how the energy conservation is imposed and how velocity rescaling is done
% 
% ^.

\textbf{Local diabatization}.
The integration of Eqn. \ref{eqn:electronic_se} in the adiabatic basis becomes numerically unstable if the non-adiabatic couplings are strongly peaked. Alternatively, the electronic Schr\"{o}dinger equation
can be transformed into a locally diabatic basis, in which the couplings become smooth functions of the nuclear displacement\cite{local_diabatization_semiempirical,local_diabatization_pyridone}. This integration scheme has been developed by Granucci and Persico and implemented in the dynamics program Newton X\cite{newton_X}. Because of its importance for weakly coupled chromophores a detailed derivation is given.

The electronic wavefunction can be expanded in a diabatic basis:
\begin{equation}
\ket{\Psi(\vec{R}(t))} = \sum_k D_k(t) \ket{\Phi_k(\vec{R}(t))}
\end{equation}
The diabatic basis $\{\Phi_k\}$ is related to the adiabatic basis $\{\Psi_i\}$ by a unitary transformation $\mathbf{T}$:
\begin{equation}
\ket{\Psi_i} = \sum_j \ket{\Phi_j} T_{ji}
\end{equation}
which transforms the expansion coefficients according to
\begin{equation}
D_i(t) = \sum_j T_{ij} C_j(t)
\end{equation}
The diabatic basis is characterized by the fact that, at least locally around some reference geometry $\vec{R}(0)$, it remains constant for displacements of the nuclear trajectory $\vec{R}(\Delta t)$:
\begin{equation}
\bracket{\Phi_i}{\frac{d}{dt} \Phi_j} = \bra{\Phi_i} \vec{\nabla}_{\vec{R}} \ket{\Phi_j} \cdot \frac{d \vec{R}}{dt} = 0 \label{eqn:diabatic_property}
\end{equation}
The reference geometry is chosen as the nuclear geometry at the beginning of a nuclear time step, $\vec{R}(t=0)$. At this reference geometry the adiabatic and diabatic bases conicide:
\begin{align}
\mathbf{T}(t=0) &= \mathbf{1} \\
\ket{\Psi_i(0)} &= \ket{\Phi_i(0)} \\
C_i(0) &= D_i(0)
\end{align}
At the end of the nuclear time step $t=\Delta t$, the adiabatic wavefunction $i$ will have evolved into a mixture of diabatic states:
\begin{equation}
\ket{\Psi_i(\Delta t)} = \sum_j \ket{\Phi_j(\Delta t)} T_{ji}(\Delta t)
\end{equation}
The overlap matrix between adiabatic states at the beginning and end of the time step can formally be written using the diabatic basis:
\begin{equation}
S_{ij}(\Delta t) = \bracket{\Psi_i(0)}{\Psi_j(\Delta t)} = \sum_k \bracket{\Phi_i(0)}{\Phi_k(\Delta t)} T_{kj}(\Delta t) \label{eqn:overlapS}
\end{equation}
Substituting a Taylor expansion of the diabatic states in Eqn. \ref{eqn:overlapS},
\begin{equation}
\ket{\Phi_j(\Delta t)} \approx \ket{\Phi_j(0)} + \ket{\frac{d \Phi_j}{dt}}\Big\vert_{t=0} \Delta t
\end{equation}
and using the equality of the adiabatic and diabatic states at $t=0$, shows that
\begin{equation}
S_{ij}(\Delta t) = \sum_{k} \left( \delta_{ik} + \underbrace{\bracket{\Phi_i}{\frac{d\Phi_k}{dt}}}_{\approx 0} \Delta t \right) T_{kj}(\Delta t).
\end{equation}
Because of the defining property of the diabatic basis in Eqn. \ref{eqn:diabatic_property}, the diabatic-to-adiabatic transformation matrix is equal to the overlap matrix:
\begin{equation}
\mathbf{T}(\Delta t) \approx \mathbf{S}(\Delta t)
\end{equation}
$\mathbf{T}$ should be exactly unitary, but $\mathbf{S}$ is not. Although the adiabatic states form an orthonormal basis of the electronic Hilbert space at the nuclear geometry $\vec{R}(t)$, i.e.
\begin{equation}
\sum_{k} \ket{\Psi_k(\vec{R}(t))}\bra{\Psi_k(\vec{R}(t))} = \mathbf{1},
\end{equation}
in practice the number of excited states has to be truncated, so that the resolution of the identity incurs a small error $\epsilon$:
\begin{equation}
\sum_{k=1}^{N_{\text{st}}} \ket{\Psi_k(\vec{R}(t))}\bra{\Psi_k(\vec{R}(t))} = \mathbf{1} + \mathbf{\epsilon}.
\end{equation}
As a consequence $\mathbf{S}(\Delta t)$ is only approximately unitary 
\begin{equation}
\begin{split}
\left( \mathbf{S}^{\dagger}(\Delta t) \mathbf{S}(\Delta t) \right)_{ij} &= \sum_k \bracket{\Psi_i(\vec{R}(t+\Delta t))}{\Psi_k(\vec{R}(t))}\bracket{\Psi_k(\vec{R}(t))}{\Psi_j(\vec{R}(t+\Delta t))} \\
 &= \delta_{ij} + \epsilon_{ij}
\end{split}
\end{equation}
To restore unitarity artificially, $\mathbf{S}(\Delta t)$ is orthogonalized by L\"{o}wdin's procedure.

We get the diabatic Hamiltonian by transforming the adiabatic hamiltonian $\mathbf{E}$, which is diagonal,
\begin{equation}
E_i \delta_{ij} = \bra{\Psi_i} \hat{H}_{\text{elec}} \ket{\Psi_j},
\end{equation}
to the diabatic basis using $\mathbf{T}(\Delta t)$:
\begin{equation}
\mathbf{H}^{\text{diab}} = \mathbf{T} \mathbf{E} \mathbf{T}^{\dagger}
\end{equation}

The electronic Schr\"{o}dinger equation for the diabatic expansion coefficients becomes particularly simple because the dynamic coupling vanishes approximately in this basis:
\begin{equation}
\imath \hbar \frac{d D_j}{dt} = \sum_i \left(H^{\text{diab}}_{ji} - \imath \hbar \cancel{\bracket{\Phi_j}{\frac{d}{dt} \Phi_i}} \right) D_i(t)
\end{equation}

The diabatic hamiltonian is interpolated between the beginning of the time step (where it agrees with the adiabatic one) and the end:
\begin{equation}
\mathbf{H}^{\text{diab}}\left(\Delta t / 2\right) = \frac{1}{2} \left( \mathbf{E}(0) + \mathbf{H}^{\text{diab}}(\Delta t) \right)
\end{equation}

The diabatic Schr\"{o}dinger equation can be integrated exactly by a matrix exponential, giving
the unitary propagator for advancing the diabatic coefficients to the end of the time step:
\begin{equation}
\mathbf{U}(\Delta t) = \exp\left(-\frac{\imath}{\hbar} \mathbf{H}^{\text{diab}}\left(\Delta t / 2\right)  \Delta t \right)
\end{equation}

The adiabatic coefficients, from which the hopping probabilities are calculated, are advanced by transforming to the diabatic basis, applying the propagator and transforming back:
\begin{equation}
\vec{C}(\Delta t) = \mathbf{T}^{\dagger}(\Delta t) \exp\left(-\frac{\imath}{\hbar} \frac{1}{2} \left[\mathbf{E}(0) + \mathbf{T}(\Delta t) \mathbf{E}(0) \mathbf{T}^{\dagger}(\Delta t) \right] \Delta t \right) \vec{C}(0)
\end{equation}

The diagonal elements of the locally diabatic hamiltonian, $H_{ii}(\vec{R}(t))$, change smoothly along the trajectory.

\begin{figure}[h!]
\includegraphics[width=0.6\textwidth]{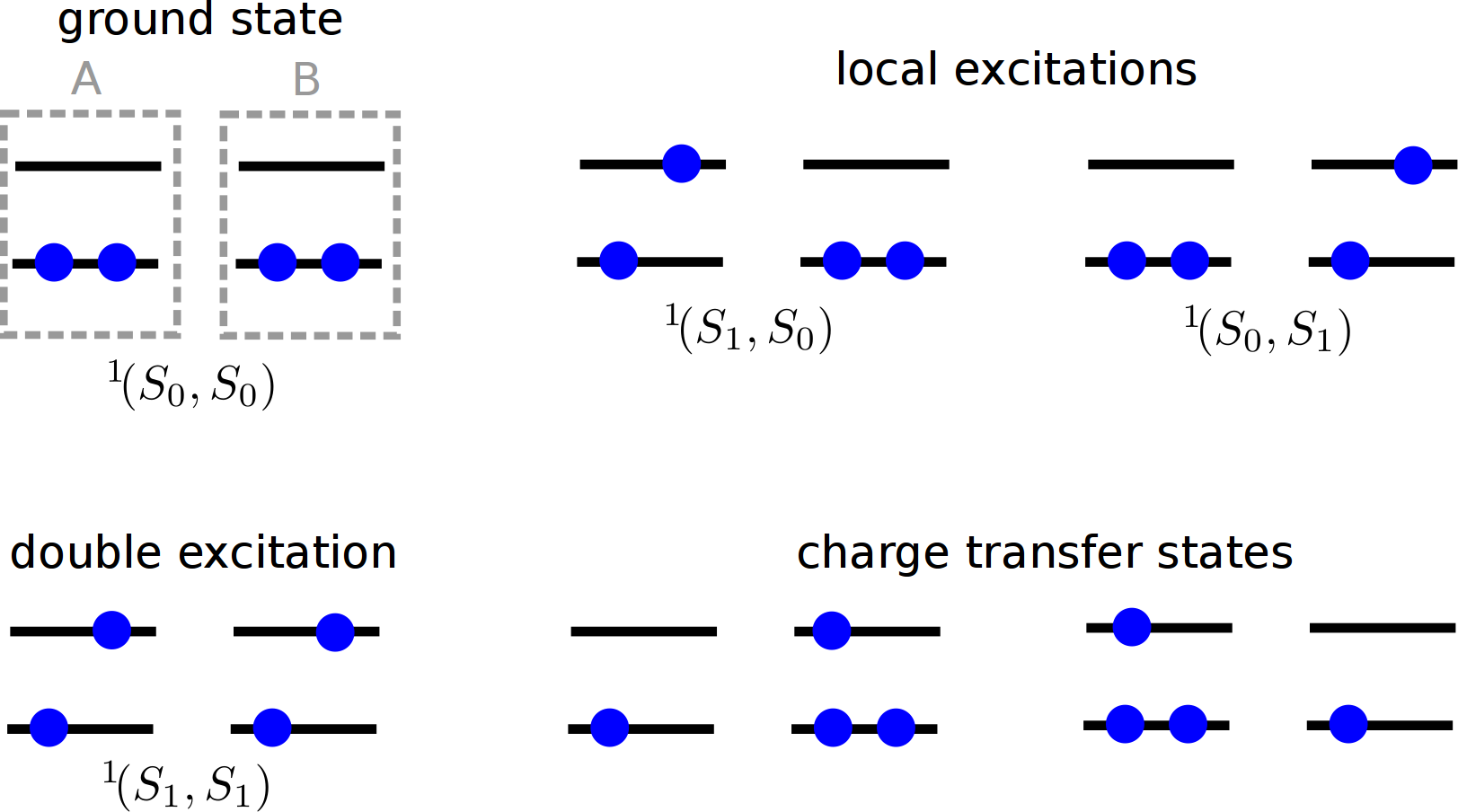}
\caption{Types of excitations for a pair of weakly interacting chromophores.}
\label{fig:dimer_states}
\end{figure}

\begin{figure}[h!]
\includegraphics[width=0.8\textwidth]{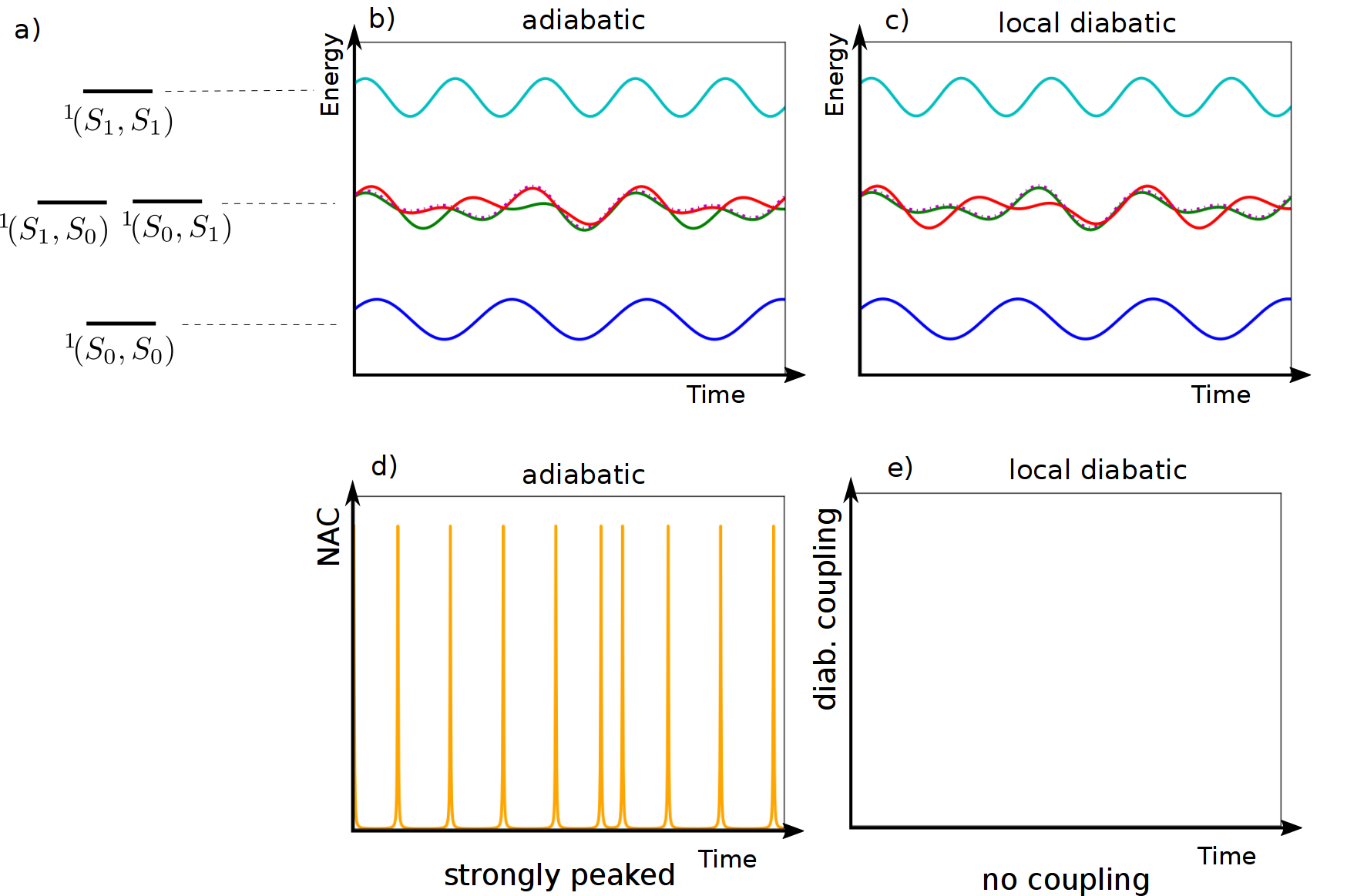}
\caption{Energies \textbf{(b,c)} and couplings \textbf{(d,e)} in the adiabatic and locally diabatic bases for a fictitious trajectory of a weakly coupled pair of molecules A and B. \textbf{a)} The blue curve shows the ground state energy, the green and red curves the energies of excitons localized on either the A or B molecule and the turquoise line a doubly excited state that would be neglected in TD-DFT. For details see main text.}
\label{fig:local_diabatization_schematic}
\end{figure}

The basic idea of local diabatization is illustrated in Figs. \ref{fig:dimer_states} and \ref{fig:local_diabatization_schematic}. In Fig. \ref{fig:dimer_states}
the type of excited states that can occur for two weakly interacting chromophores $A$ and $B$ are depicted schematically: The localized excitations on each monomer are energetically close and can hybridize to form a pair of delocalized exciton states. 
The ordering of the exciton states depends on the geometric arrangement of the chromophores: 
In a head to tail arrangement of the transition dipoles the bright state is lowered in energy
($ E(\longrightarrow\longrightarrow) < E(\longrightarrow\longleftarrow)$, J-aggregate), 
while in a parallel arrangement the dark state is stabilized 
($E({\longleftarrow \atop \longrightarrow}) < E({\longrightarrow \atop \longrightarrow})$, H-aggregate). 
(The ordering is easy to remember when one considers the attraction or repulsion between little bar magnets (``$\rightarrow$'' = ``$+-$'') instead of transition dipoles.)
%If in turn the excitation is localized to a single molecule, it is called a Frenkel exciton.
% * <roland.mitric@uni-wuerzburg.de> 2017-03-05T11:26:59.855Z:
% 
% This is not precise. Frenkel excitons may be delocalised over several molecules. If the exciton is localised on a scale comparable to a lattice constant then it is called Frenkel exciton, in contrast to Wannier excitons which are delocalised over "many" unit cels. 
% 
% ^.
Double excitations are usually higher in energy and cannot be described with linear response TD-DFT, anyway. 
The lowest adiabatic states are usually a superposition of the exciton states with some fraction of charge transfer character, that varies with the nuclear geometry. 

Fig. \ref{fig:local_diabatization_schematic} shows adiabatic and local diabatic energies along a fictitious trajectory for two completely uncoupled chromophores. The dashed line marks the current electronic state. The locally excited states $^1(S_1,S_0)$ and $^1(S_0,S_1)$ are not coupled at all. The spikes in the non-adiabatic coupling \textbf{(d)} are artifacts of the adiabatic representation, since adiabatic state labels need to switch each time two electronic energy levels cross. After transforming to a locally diabatic basis, the coupling is eliminated \textbf{(e)}. The local diabatic energies \textbf{(c)} smoothly connect energy levels at neighbouring time steps. The plot in \textbf{(c)} demonstrates clearly that the character of the electronic state never changes despite the frequent surface hops in \textbf{(b)}.

\FloatBarrier

\textbf{Conical intersections with $S_0$.} 
 Tight-binding TD-DFT inherits many problems from full TD-DFT. One of them is the absence of conical intersections between the ground state and any excited state\cite{levine_tddft_conical_intersections}.
There are two conditions for a conical intersection between two electronic states:
\begin{itemize}
  \item The energies of the two states have to be degenerate and
  \item the coupling between the states has to vanish.
\end{itemize}
In the space of $N$ internal degrees of freedom the points where these two conditions are satisfied form a $N-2$ dimensional surface called the intersection seam. If the potential energies of the two states are plotted around the conical intersection in two directions perpendicular to this surface, the potential energy surfaces have the characteristic form of a double conus.
 Since linear-response TD-DFT lacks double excitations the coupling between the ground state and all excited states vanishes due to Brillouin's theorem independently of the nuclear coordinates. Therefore the number of conditions
 is reduced and the intersection seam between the ground state and an excited state has the wrong dimensionality $N-1$\cite{levine_tddft_conical_intersections}. Movement along the non-adiabatic coupling vector does not lift the degeneracy and the potential energy surfaces have the shape of two intersecting planes instead of a conus. 
 
In our implementation of surface hopping the wrong topology of the intersection seam is mitigated by giving special treatment to surface hops to the ground state: 
\begin{itemize}
 \item If the energy gap to the ground state falls below a threshold, a hop to the ground state is forced irrespective of the quantum populations.
 \item After reaching the ground state, jumps back to higher states are suppressed and the trajectory continues on $S_0$.
\end{itemize}

Surface hopping trajectories are not very sensitive to the topology of the intersection seam, since each trajectory explores the potential energy surface only along a one dimensional path and the effects of Berry phases on the nuclear wavepacket are neglected. A study on oxirane \cite{casida_oxirane} showed that reasonable photochemical reaction paths are predicted with TD-DFT in combination with surface hopping despite the absence of true conical intersections to the ground state.
%%%%%%%%%%%%%%%%%%%%%%%%%%%%%%%%%%%%%%%%%%%%%%%%%%%%%%%%%%%%%%%%%%%%%%%%

\section{Results}
\subsection{Poly(fluorenemethylene)}
\label{sec:fluorene}
Fluorene is an aromatic hydrocarbon that owes its name to its fluorescence in the ultraviolet spectral region. 
Recently Rathore\cite{rathore_fluorene_synthesis} and coworkers synthesized $\pi$-stacked arrays of polyfluorenes (see Fig. \ref{fig:F5_minima_S0-S1}). The linker atoms keep the cofacial chromophores in close van der Waals contact and guarantee the electronic coupling between neighbouring units. The existence of this coupling has been deduced experimentally from the decrease of the ionization and oxidation potentials with increasing number of fluorene units.
The methodology developed in the present contribution allows us to explore the excited state relaxation and excimer formation dynamics in large multichromophoric systems. 
Excimer formation is an undesirable effect in light harvesting devices or organic electronics, since it traps an excitation and stops the coherent propagation of a delocalized excitations. The formation of excimers is accompanied by a geometric distortion as a pair of molecules move closer to each other and align. In $\pi$-stacked polyfluorene the eclipsed conformation, where the chromophore units are perfectly aligned, is not the one with lowest energy. Instead neighbouring units are slightly rotated around the axis of the polymer chain (see Fig. \ref{fig:F2_angle_scan_composite}). The authors of Ref. \cite{rathore_fluorene_theory} showed that an excitation can only be delocalized over many fluorene units, if they are all in the eclipsed conformation. On the other hand this perfect $\pi$-stacking favours the formation of excimer pairs. As a result excitons become localized to pairs of fluorene excimers and propagate by a hopping mechanism. 

These conclusion were made based on TD-DFT calculations at stationary points of the ground and first excited state. However, exciton propagation and excimer formation are dynamic processes. It would therefore be interesting to exploit the speed of tight-binding DFT (DFTB) to investigate the dynamics directly through simulation. Since tight-binding DFT is not as reliable as full DFT, in a first step one has to check that it can reproduce the results from the previous TD-DFT study. 

\subsubsection{Fluorene monomer \textbf{F1}}
\begin{figure}[h!]
\includegraphics[width=0.7\textwidth]{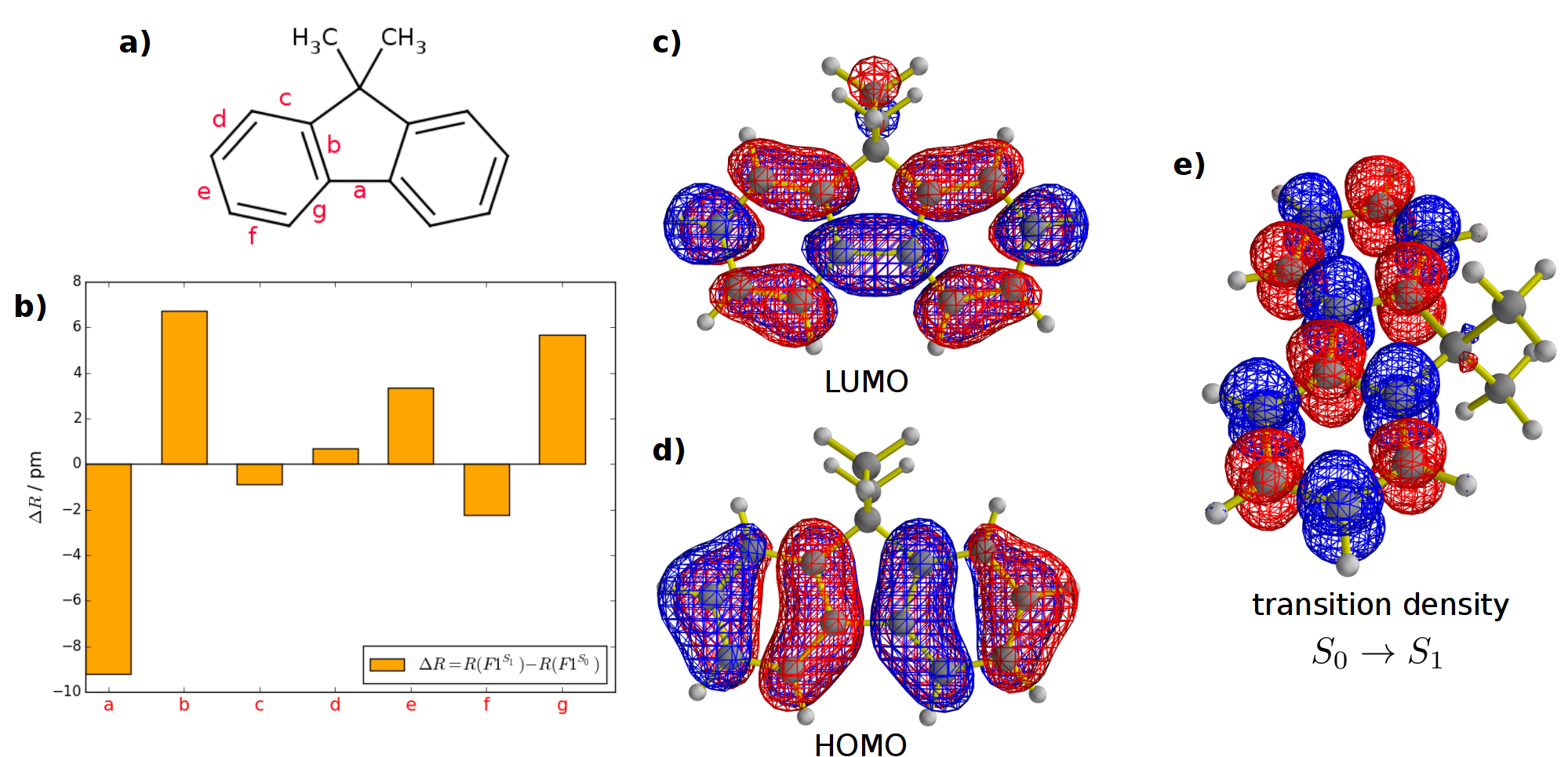}
\caption{Changes of bond lengths between minima on $S_0$ and $S_1$ as calculated with DFTB. \textbf{a)} Names of C-C bonds, \textbf{b)} bar plot showing differences. The $S_1$ state consists of a HOMO to LUMO excitation, DFTB orbitals are shown in \textbf{c)} and \textbf{d)}, the transition density in \textbf{e)}.}
\label{fig:F1_bond_length_changes_S1-S0}
\end{figure}

First the monomer unit, 9,9-dimethyl-fluorene, is optimized using tight-binding DFT (which includes a long-range correction and a dispersion correction) on the $S_0$ and $S_1$ states. The HOMO and LUMO and the differences of bond lengths between the minima of $S_0$ and $S_1$ are depicted in Fig. \ref{fig:F1_bond_length_changes_S1-S0}. The same pattern of changes in bond lengths is observed as in full DFT: bond lengths where the HOMO is antibonding and the LUMO is bonding are shortened while bonds where the LUMO is antibonding and the HOMO is bonding expand. 

The vertical excitation energy of $E_{\text{exc}} = 4.22$ eV obtained with DFTB is a little bit too low as compared to the B1LYP-40/6-31G(d) result of 4.89 eV. Although the relaxation on $S_1$ leads only to minor changes in the geometry, the emission energy is red-shifted by 0.88 eV relative to the excitation energy, $E_{\text{em}} = 3.34$ eV, while the full DFT calculation predict an emission energy of $4.21$ eV and consequently a red-shift of $0.68$ eV.

\subsubsection{$\pi$-stacked fluorene dimer \textbf{F2}}

\begin{figure}[h!]
\includegraphics[width=0.7\textwidth]{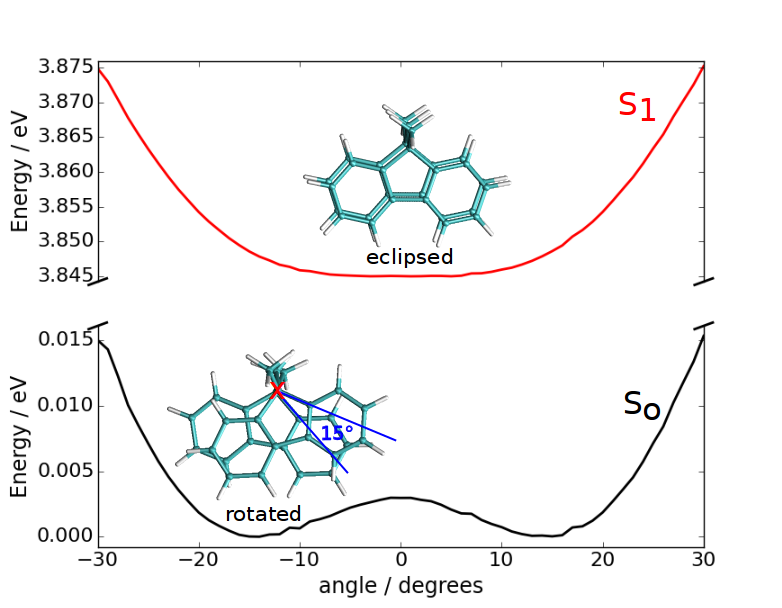}
\caption{Relaxed scan of the potential energy curve for rotation around the line passing perpendicularly through the point marked with a red x. The geometries were relaxed using the AM1\cite{am1} method in Gaussian\cite{g09}, the $S_0$ and $S_1$ energies were then computed using DFTB. In the ground state the fluorene units are rotated by $21^{\circ}$ (DFTB) or $15^{\circ}$ (AM1), while in the $S_1$ the eclipsed geometry is preferred.}
\label{fig:F2_angle_scan_composite}
\end{figure}

For the fluorene dimer initially two structures are considered, the one in which a fluorene molecule eclipses the other and the second one, where they are rotated by an angle of approximately 21$^{\circ}$. Optimization with DFTB on the ground state shows that both of them are local minima, but the rotated structure is 0.1 eV more stable than the eclipsed one. In turn, on the $S_1$ surface the global minimum appears to be located at the eclipsed geometry which is 0.1 eV more stable. Fig. \ref{fig:F2_angle_scan_composite} shows a relaxed scan along the angle between the two fluorene units. 

\begin{figure}[h!]
\includegraphics[width=0.7\textwidth]{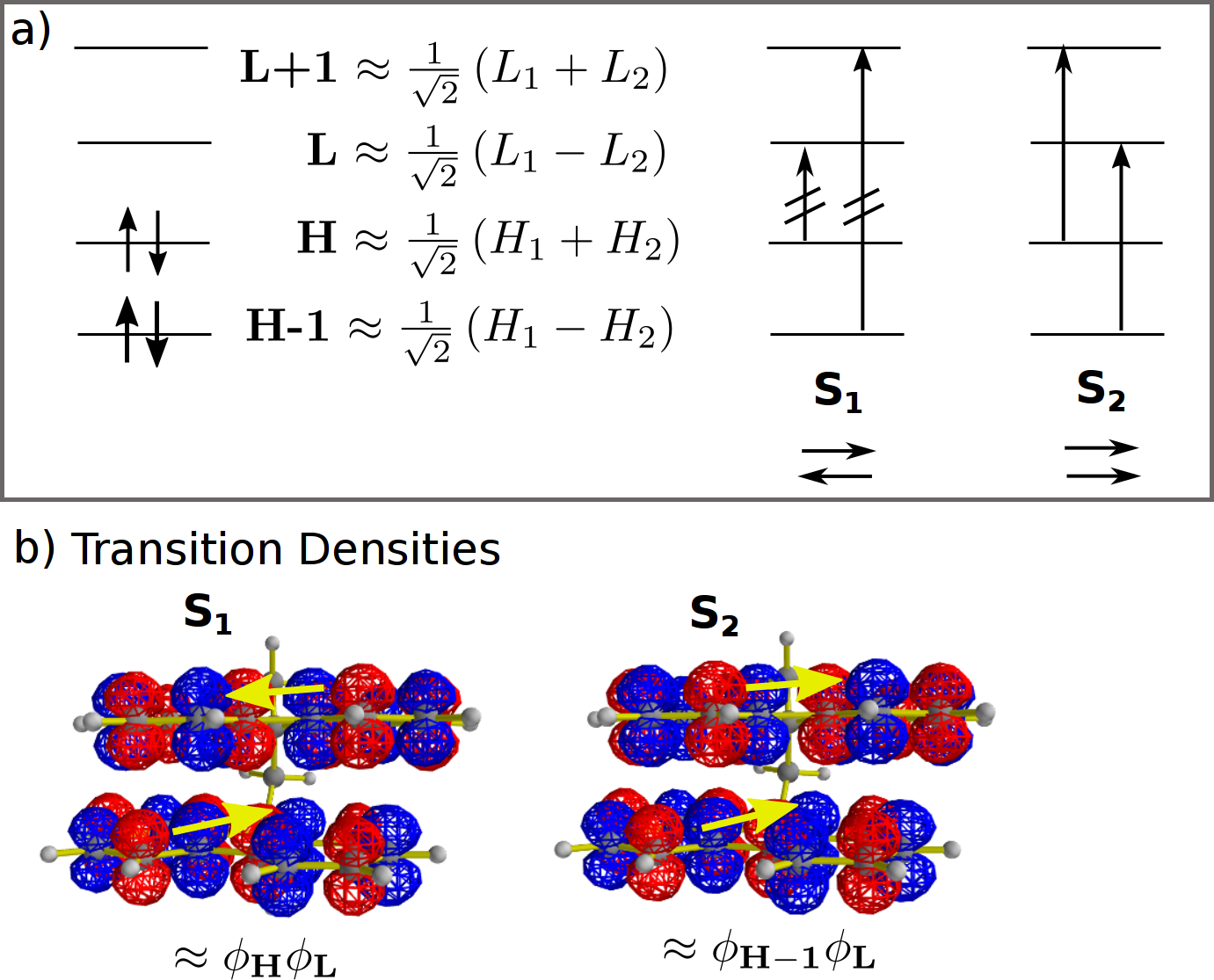}
\caption{\textbf{F2}. \textbf{a)} Frontier orbitals in terms of monomer orbitals and possible excitations. \textbf{b)} Transition densities with monomer transition dipoles superimposed.}
\label{fig:F2_frontier_orbitals}
\end{figure}

The frontier orbitals of the dimer can be approximately constructed as linear combination of the monomer HOMO and LUMO orbitals. 
The energetic order of the orbitals is obvious from the condition that the energy should increases with the number of nodes.
The orbital combinations and possible singlet transitions are sketched in Fig. \ref{fig:F2_frontier_orbitals}. 
Since only excitation of (approximately) the same symmetry can mix, the lowest two excited state have to contain the orbital transitions
shown in Fig. \ref{fig:F2_frontier_orbitals}a). The approximate expressions are:

\begin{eqnarray*}
\ket{S_1} &\approx& \frac{1}{\sqrt{2}} \left[ \ ^1\!(H-1 \to L+1) - ^1\!(H \to L) \right] \quad \text{(dark)} \\
\ket{S_2} &\approx& \frac{1}{\sqrt{2}} \left[ \ ^1\!(H \to L+1)   - ^1\!(H-1 \to L)\right] \quad \text{(bright)}
\end{eqnarray*}

The lowest excited state $S_1$ is dark since in the HOMO-LUMO transition the monomer dipole moments would point in opposite direction canceling each other (see Fig. \ref{fig:F2_frontier_orbitals}\textbf{b}).
The $S_2$ is the bright state since the monomer transition dipoles are parallel.

\begin{figure}[h!]
\includegraphics[width=0.7\textwidth]{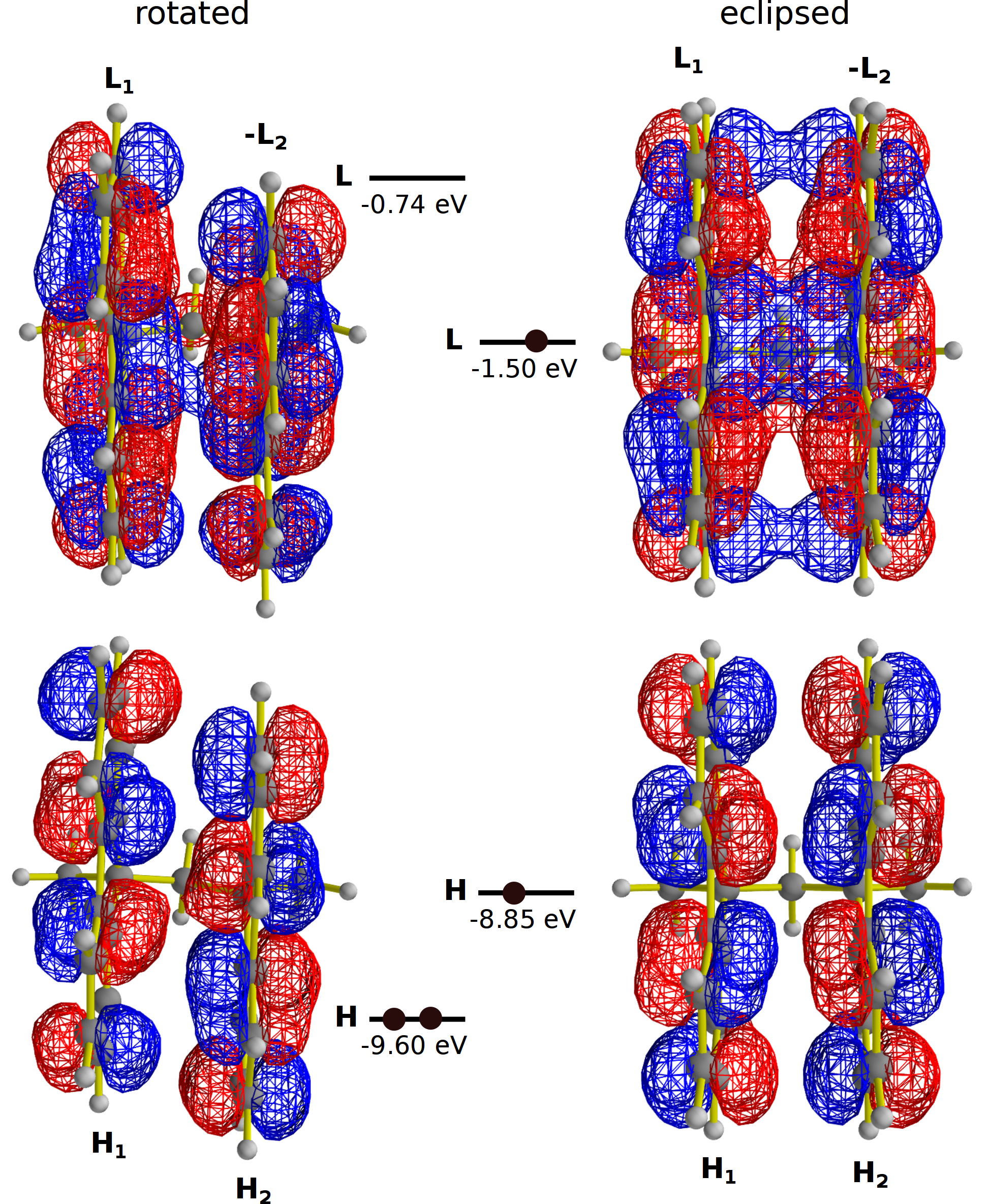}
\caption{DFTB frontier orbitals for the rotated and the eclipsed local minima of \textbf{F2}. The rotated geometry (left) is more stable in the ground state because of the attractive overlap between $H_1$ and $H_2$ in the HOMO. In the eclipsed geometry (right) the energy of the LUMO is lowered stabilizing the $S_1$ state.}
\label{fig:F2_rot_eclipsed_stabilization}
\end{figure}

Fig. \ref{fig:F2_rot_eclipsed_stabilization} compares the orbital interactions in the rotated and eclipsed conformations. In the ground state the two fluorene units are rotated slightly because this maximizes the constructive overlap between the monomer orbital $H_1$ and $H_2$ and thus stabilizes the HOMO. 
The LUMO is destabilized in the rotated conformation. In the $S_1$ excited state the eclipsed conformation is preferred, since it favors the overlap between $L_1$ and $-L_2$ and stabilizes the LUMO, which is occupied in $S_1$.

\FloatBarrier
\subsubsection{Oligomers \textbf{F3}-\textbf{F5}}

\begin{figure}[h!]
\includegraphics[width=0.8\textwidth]{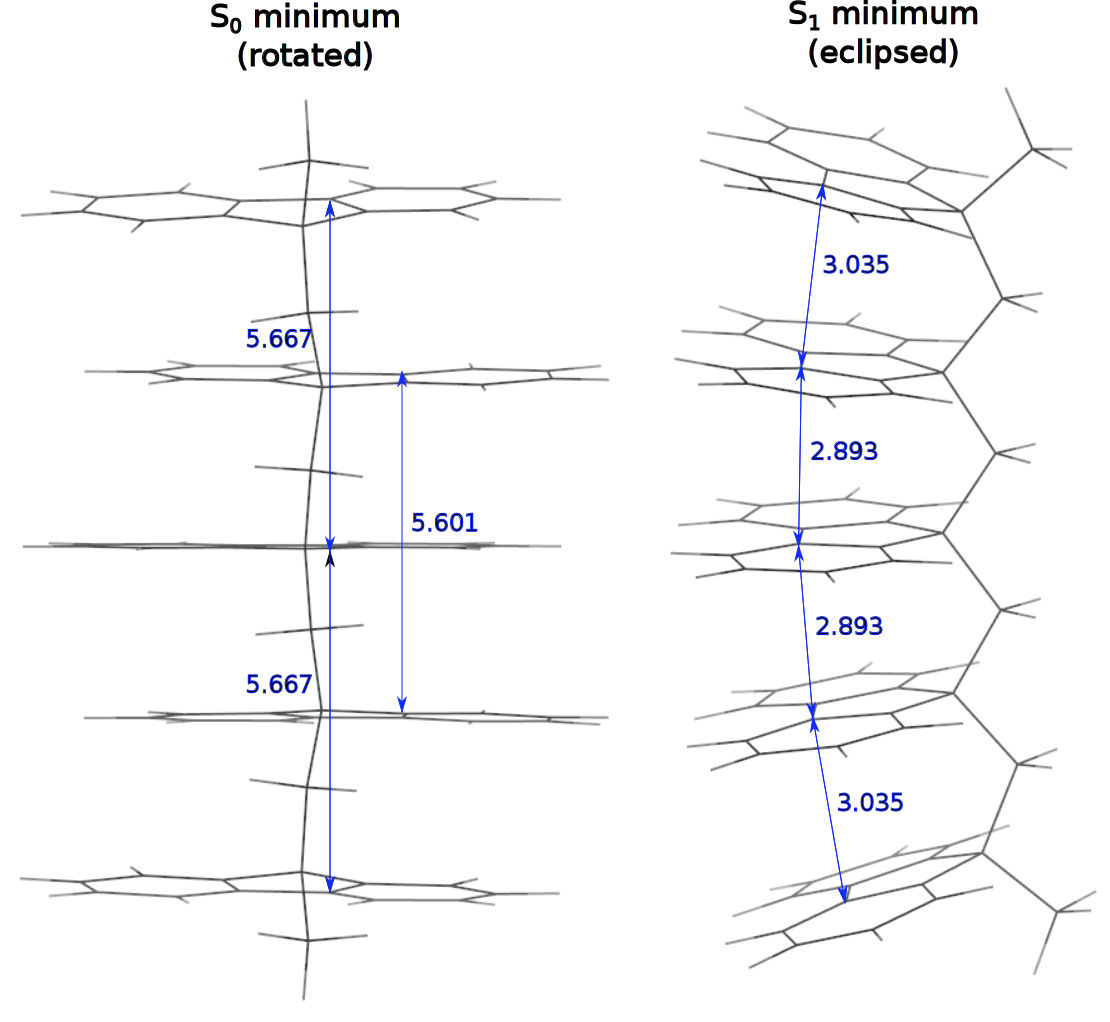}
\caption{Optimized pentamer (\textbf{F5}) geometries on $S_0$ and $S_1$. Distances are in $\AA$.}
\label{fig:F5_minima_S0-S1}
\end{figure}

The oligomers \textbf{F3}-\textbf{F5} were also optimized using tight-binding DFT on the ground and first excited states. In all of them the most stable conformation is the rotated one on $S_0$ and the eclipsed one on $S_1$. The optimized geometry for \textbf{F5} is shown in Fig. \ref{fig:F5_minima_S0-S1}.

\begin{figure}[h!]
\includegraphics[width=0.4\textwidth]{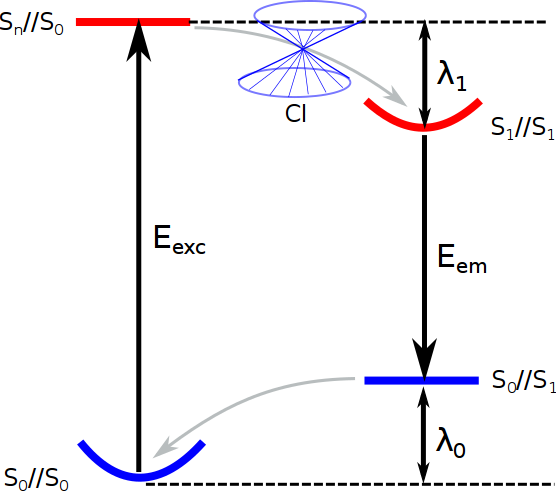}
\caption{Energetics of excimer formation. Shown are the vertical excitation energy to the brightest state $E_{\text{exc}}$, the emission energy $E_{\text{em}}$ and the structural reorganization energies in the ground and excited states $\lambda_0$ and $\lambda_1$ (adapted from \cite{rathore_fluorene_theory}).}
\label{fig:fluorene_relaxation_pes_sketch}
\end{figure}

In the $\pi$-stacked fluorene oligomers the $S_1$ will not necessarily be the brightest electronic state. 
The excited state, where all monomer transition dipoles are parallel will have the highest oscillator strength, in \textbf{F2} this is the 2nd, in \textbf{F3} the 3rd, in \textbf{F4} the 4th and in \textbf{F5} the 5th excited states. 
No matter which higher state $S_n$ is excited, ultimately the lowest excited state will be populated through non-adiabatic relaxation. The $S_1$ state is separated through a large energy gap from the ground state and can only decay through emission of a photon of energy $E_{\text{em}}$ (fluorescence). Therefore the emission spectrum can be calculated from the energy of the lowest accessible minimum on the $S_1$ surface. This is illustrated in Fig. \ref{fig:fluorene_relaxation_pes_sketch}: After vertical excitation with energy $E_{\text{exc}}$ to the bright state $S_n$ the wavepacket can undergo non-adiabatic transitions through conical intersections (CI) until it reaches the long-lived first excited state $S_1$. The excess energy $\lambda_1$ effects the relaxation in the individual fluorene units and the change of their orientation and will be ultimately dissipated to the solvent. 

Experimentally\cite{rathore_fluorene_emission} the lowest peak in the absorption spectrum is observed at 302 nm which shifts to 305 nm in the dimer. In the emission spectrum the lowest momoner peak is found at 315 nm and shifts to 394 nm for the dimer. Absorption and emission spectra of trimer, tetramer, pentamer and hexamer are almost indistinguishable from the dimer spectra with tiny red shifts. Also, no exciton splitting was observed. 
This suggests that the delocalization of an excitation is limited to two fluorene units.
In table \ref{tbl:Fn_Eexc_Eem} the theoretical and experimental emission and absorption lines are compared. In agreement with experiment, the largest red shift is seen between the monomer and the dimer, while there is much less variation in the excitation (to $S_1$) and emission energies between the dimer and the longer oligomers \textbf{F3}-\textbf{F5}. 

\begin{table}
\begin{tabular}{c|ccccc|cc}
   & $E_{\text{exc}}$ & $E_{\text{em}}$ & $\lambda_0$ & $\lambda_1$ & n    & $E_{\text{exc}}$(exp.) & $E_{\text{em}}$(exp.) \\ \hline
F1 & 4.22             & 3.34            & 0.44        & 0.44        & 1  & 4.11 & 3.94 \\
F2 & 4.22 (4.15)      & 3.06            & 0.57        & 0.58        & 2  & 4.07 & 3.15 \\
F3 & 4.35 (4.19)      & 3.08            & 0.63        & 0.63        & 3  & 4.05 & 3.14 \\
F4 & 4.34 (4.17)      & 3.07            & 0.74        & 0.37        & 4  & 4.05 & 3.14 \\
F5 & 4.34 (4.16)      & 3.02            & 0.84        & 0.48        & 5  & 4.05 & 3.14   
\end{tabular}
\caption{\textbf{Absorption and fluorescence energies.} Theoretical vertical excitation energies $E_{\text{exc}} = E(S_n // S_0) - E(S_0 // S_0)$ of the bright state $S_n$ (in brackets excitation energies of $S_1$), emission energies $E_{\text{em}} = E(S_1 // S_1) - E(S_0 // S_1)$ and reorganization energies $\lambda_0 = E(S_0 // S_1) - E(S_0 // S_0)$ and $\lambda_1 = E(S_n // S_0) - E(S_1 // S_1)$ where $E(S_n // S_0)$ means the total energy of the $n$-th excited state at the minimum geometry on $S_0$. Experimental excitation and emission energies at the peak maxima are taken from Ref.\cite{rathore_fluorene_emission}. All energies in eV.}
\label{tbl:Fn_Eexc_Eem}
\end{table}

\FloatBarrier
\subsubsection{Nonadiabatic dynamics simulations}
\begin{figure}[h!]
\includegraphics[width=1.0\textwidth]{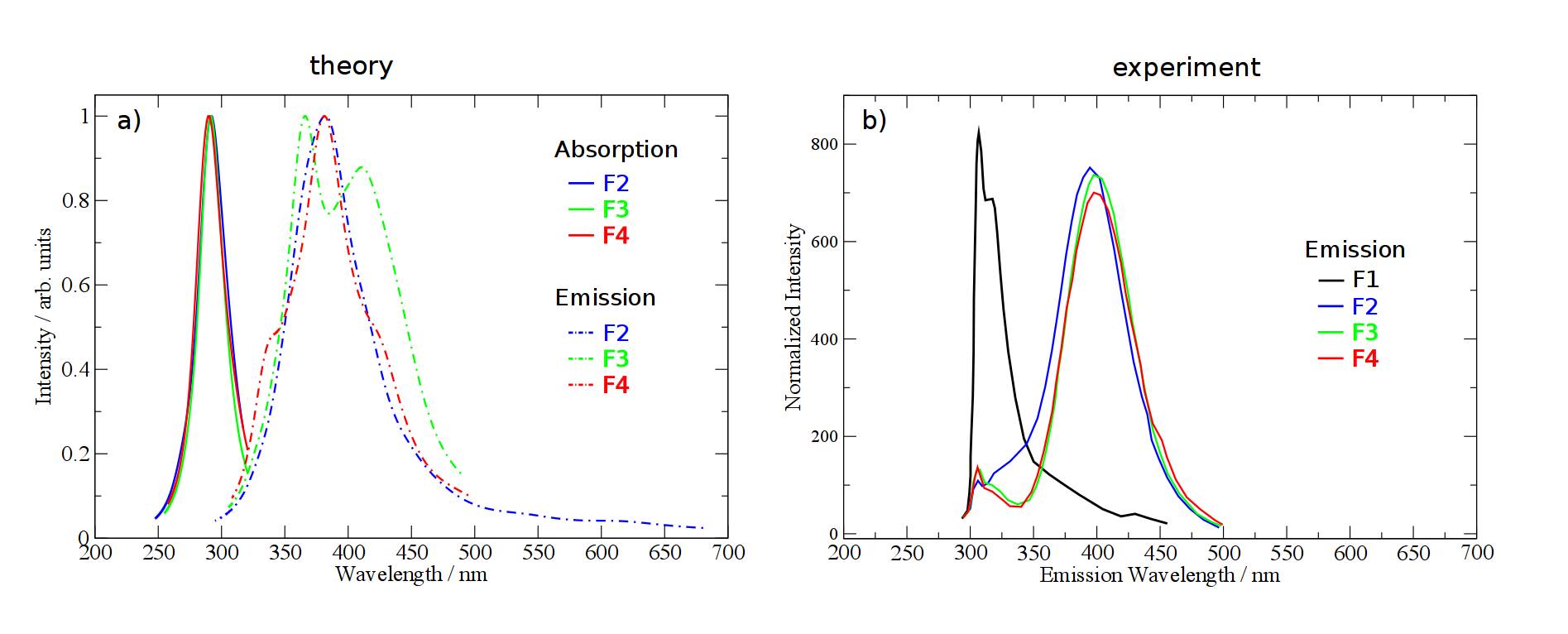}
\caption{a) Simulated absorption and emission spectra for \textbf{F2},\textbf{F3} and \textbf{F4}. The stick spectra were convolved with a Gaussian of FWHM=0.01 Hartree. b) Experimental emission spectra for $\lambda_{\text{exc}}=280$ nm were digitized from Fig. 1C in Ref.\cite{rathore_fluorene_theory}. }
\label{fig:F2_F3_F4_avg_absorption_emission}
\end{figure}

% DIMER F2
Non-adiabatic dynamics simulation are performed for the dimer \textbf{F2}, trimer \textbf{F3} and tetramer \textbf{F4}. 
The geometry is minimized on the ground state starting from the rotated conformation. 
Subsequently the Hessian matrix is computed by numerical differentiation of the analytic gradients. The Wigner distribution\cite{wigner_distribution} in the harmonic approximation is 
constructed from the normal mode displacements and frequencies. 100 initial conditions for the initial positions and momenta are sampled at random from the Wigner
distribution\cite{barbatti_initial_condition_sampling}. A well-known shortcoming of this approach is that hydrogen atoms have too large velocities\cite{klaffki_quantum_vs_thermal_sampling}. Therefore the trajectories are propagated for 1 ps on the
ground state at a constant temperature of $T=150 $K, that is controlled using a Berendsen thermostat\cite{berendsen_thermostat}, to arrive at an equilibrated distribution. For each equilibrated geometry a TD-DFTB calculation is performed and
the stick spectra from different trajectories are combined and convolved with a Gaussian function to simulate a temperature broadened absorption spectrum (see Fig. \ref{fig:F2_F3_F4_avg_absorption_emission}).

The trajectories are lifted vertically to the brightest excited state and are allowed to evolve again for 1 ps at constant energy. Non-adiabatic transitions between different
electronic states are accounted for by surface hopping. The energies of the $S_1$ state at the end of the non-adiabatic simulation from different trajectories are combined to obtain a 
theoretical fluorescence spectrum (see Fig. \ref{fig:F2_F3_F4_avg_absorption_emission}). 

\begin{figure}[h!]
\includegraphics[width=0.7\textwidth]{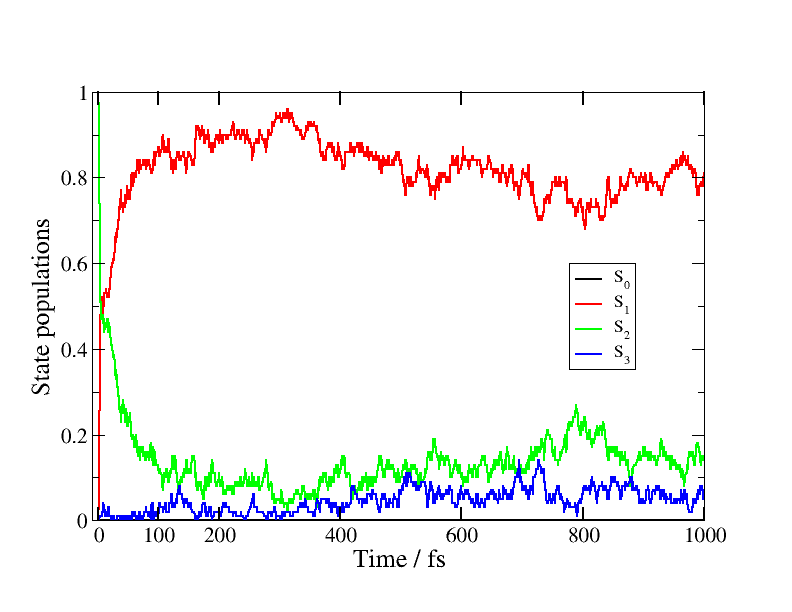}
\caption{\textbf{Dimer F2.} Adiabatic state populations averaged over 100 trajectories.}
\label{fig:F2_nonadiabatic_S2_populations}
\end{figure}

\begin{figure}[h!]
\includegraphics[width=0.7\textwidth]{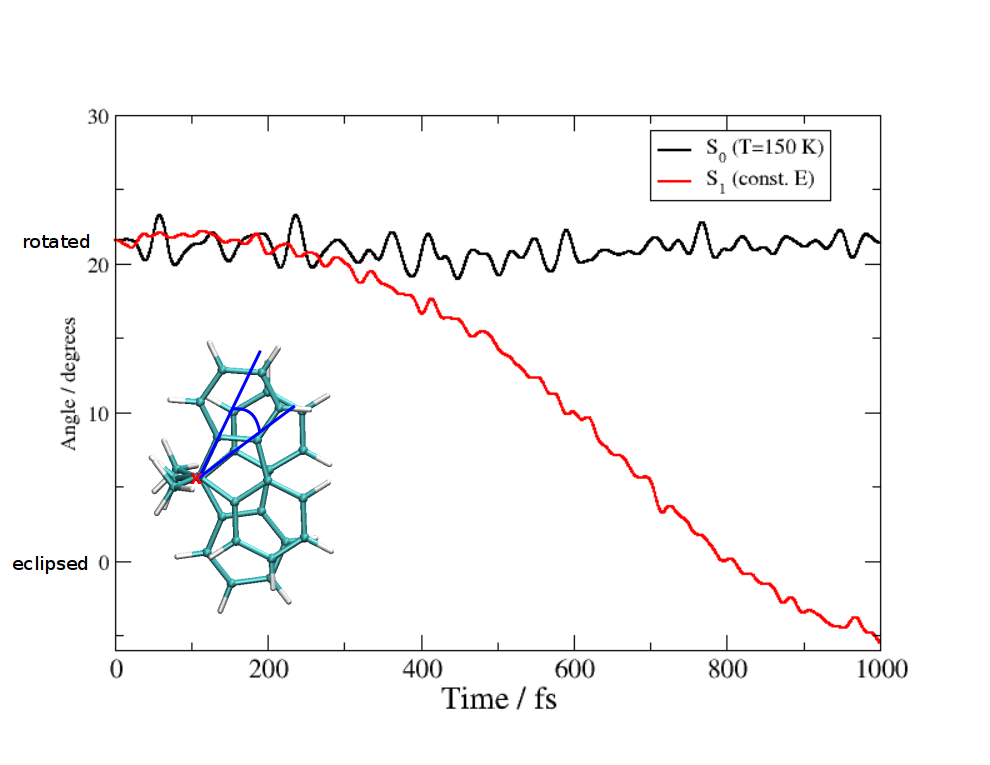}
\caption{\textbf{Dimer F2.} Dihedral angle (see inset) averaged over 100 trajectories. In the ground state the geometry oscillates around the rotated structure (black curve). After vertical excitation to $S_2$ and ultrafast non-adiabatic transition to $S_1$, the eclipsed geometry is reached after 1 ps (red curve).}
\label{fig:F2_dihedral_timeseries}
\end{figure}

\begin{figure}[h!]
\includegraphics[width=0.9\textwidth]{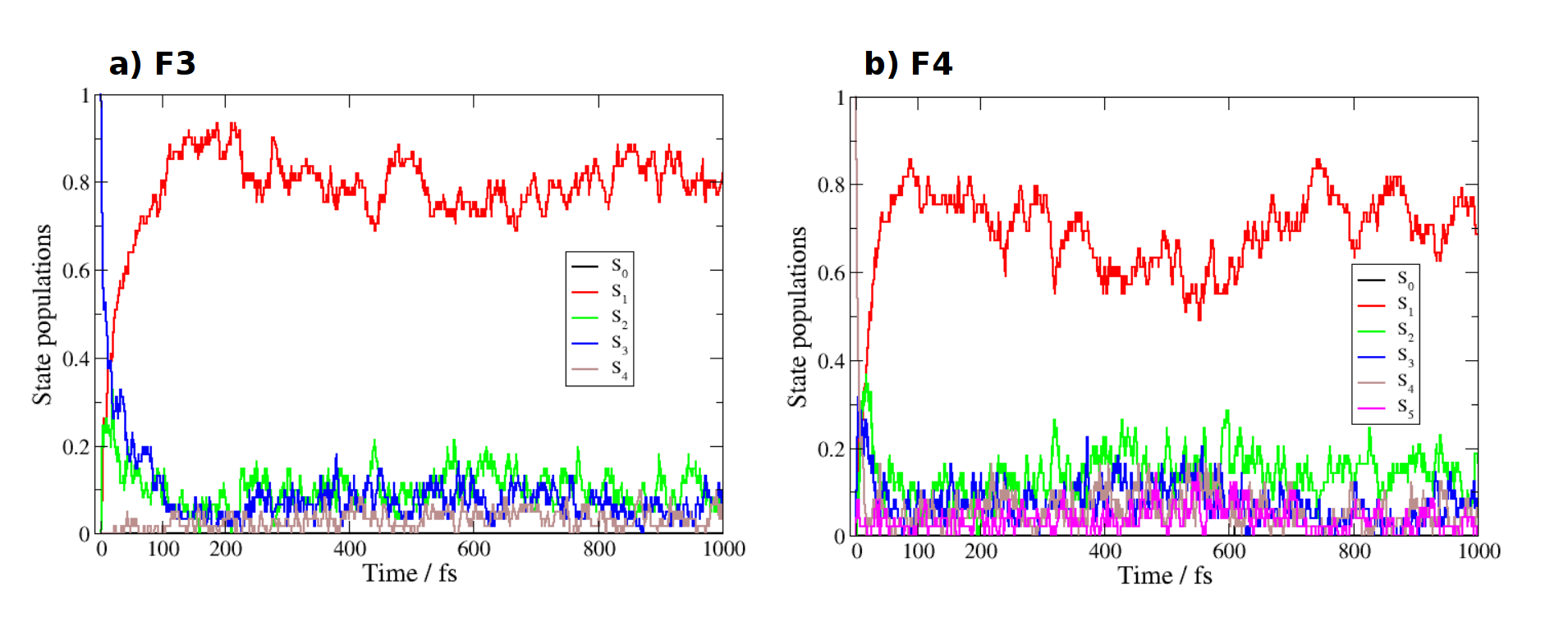}
\caption{Adiabatic state populations averaged over 50 trajectories for the \textbf{a)} trimer \textbf{F3} and \textbf{b)} tetramer \textbf{F4}.}
\label{fig:F3_F4_state_populations}
\end{figure}

\begin{figure}[h!]
\includegraphics[width=0.8\textwidth]{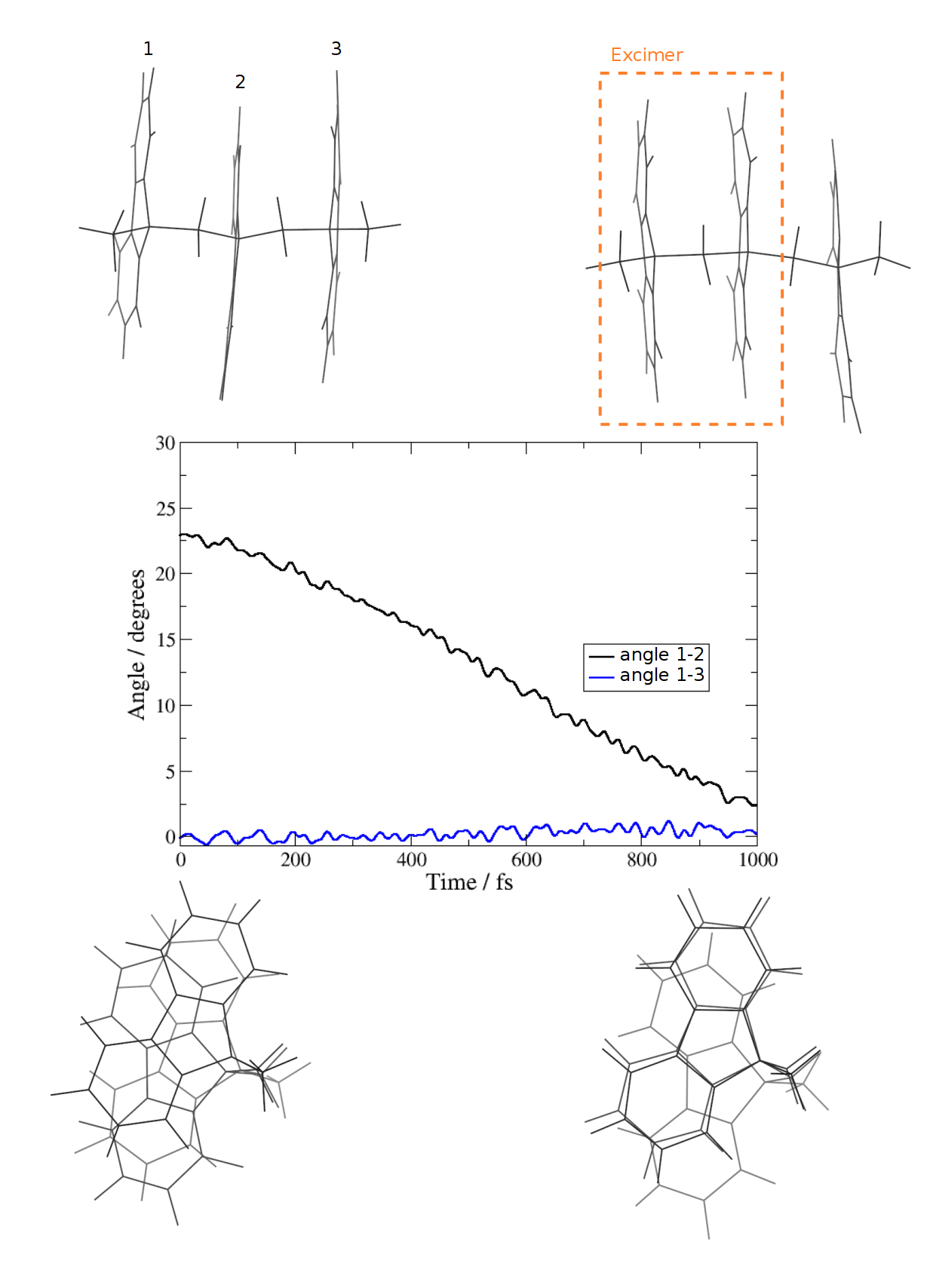}
\caption{Trimer \textbf{F3}. Dihedral angles between fluorene units 1-2 and 1-3 averaged over 50 trajectories. After 1 ps the fluorene units 1-2 are aligned in the eclipsed conformation forming an excimer.}
\label{fig:F3_excimer_formation}
\end{figure}

\begin{figure}[h!]
\includegraphics[width=0.8\textwidth]{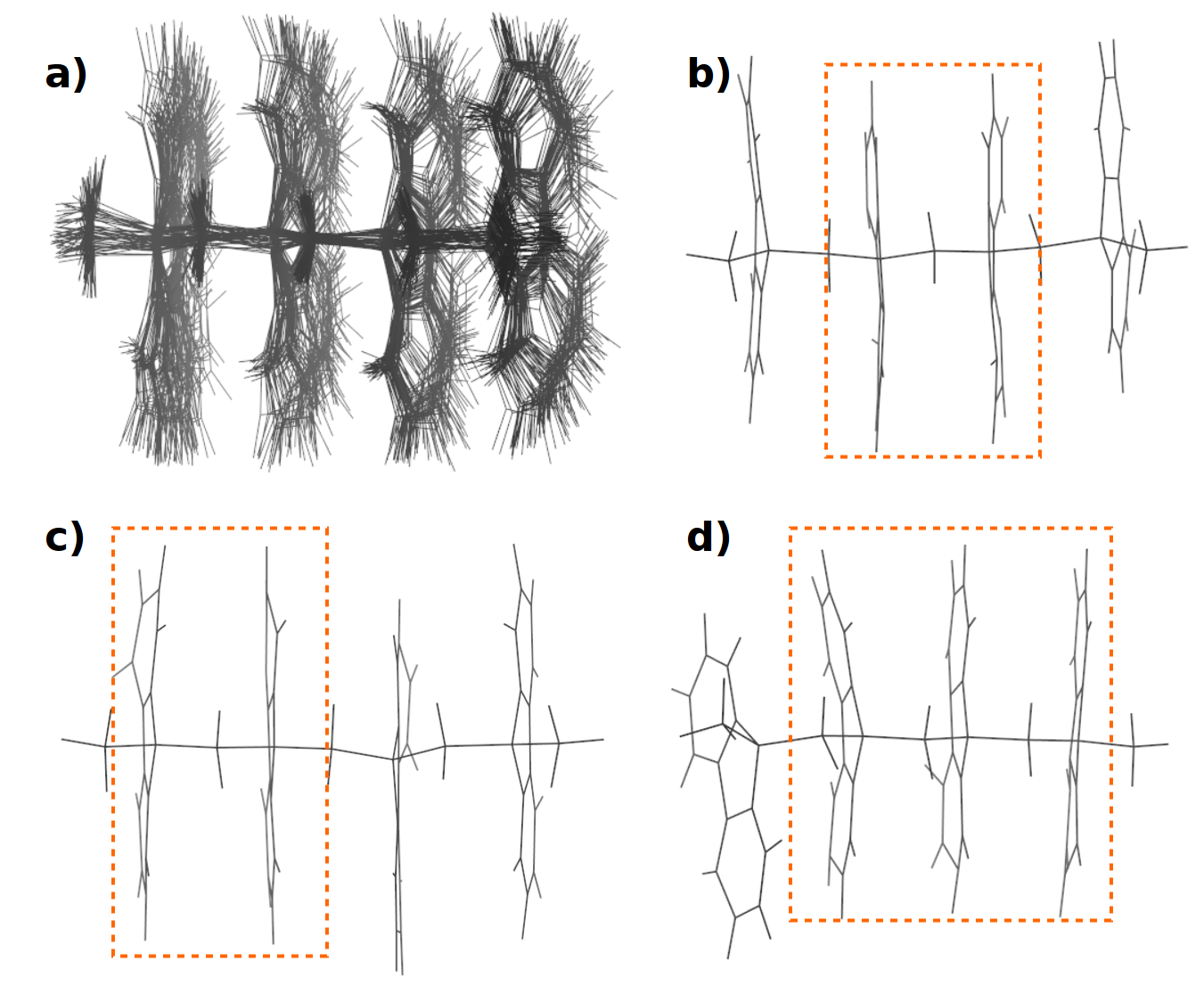}
\caption{Tetramer \textbf{F4}. \textbf{a)} superposition of all 50 trajectories at the last time step (1 ps). \textbf{b)}, \textbf{c)} and \textbf{d)} Geometries of different trajectories after 1 ps with the fluorene units that form an excited dimer or timer marked by an orange box.}
\label{fig:F4_excimer_pairs}
\end{figure}

The excited dimer \textbf{F2} decays in less than 100 fs non-radiatively (see Fig. \ref{fig:F2_nonadiabatic_S2_populations}) to the $S_1$ state and rotates slowly towards the eclipsed conformation as evidenced by a plot of the angle between the monomer units against time in Fig. \ref{fig:F2_dihedral_timeseries}. In the trimer \textbf{F3} and tetramer \textbf{F4} the bright states are $S_3$ and $S_4$, which also decay in less than 100 fs to the long-lived $S_1$ state (see Fig. \ref{fig:F3_F4_state_populations}). 
In the trimer one can observe how the excitation localizes on two of the fluorene units which rotate towards each other to form an excimer, while the 3rd unit is unaffected (see Fig. \ref{fig:F3_excimer_formation}). 
In the tetramer the excitation can localize on any of the three fluorene pairs, 1-2, 2-3 or 3-4, and depending on the initial conditions all three cases can be observed among the ensemble of trajectories (see Fig. \ref{fig:F4_excimer_pairs}). 
The quantitative agreement between the simulated absorption and emission spectra shown in Fig. \ref{fig:F2_F3_F4_avg_absorption_emission} with the experimental spectra published in Ref. \cite{rathore_fluorene_theory} suggests that the trajectories move on reasonable potential energy surfaces and reach the correct $S_1$ minimum, from which the fluorescence is observed experimentally. In addition the dynamics simulation provides the time-scale for the excimer formation: The rotation into the eclipsed conformation in \textbf{F2}, \textbf{F3} and \textbf{F4} lasts approximately 1 ps, which is a lower bound for the formation time, since one expects a damped oscillation around the new minimum. 
Since most of the trajectories for \textbf{F4} show the formation of aligned pairs of fluorene units, one can expect that in longer chains, too, the initially delocalized excitation will be trapped by the alignment of neighbouring units. 

%absorption and emission spectra agree quantitatively with the 
%long trajectories needed

\FloatBarrier
% TRIMER F3 and TETRAMER F4

\section{Conclusion}
We have given full details of the theoretical methods that are implemented in the code for 
non-adiabatic molecular dynamics simulations in the framework of tight-binding lc-TD-DFT. 
Special attention has been paid to peculiarities of weakly coupled aggregates: 
(1) Erroneous long-range charge transfer is fixed by including exact exchange for large distances 
and (2) spikes in the non-adiabatic couplings are removed by a transformation to a local diabatic basis. 

As a test example, we have calculated excited state lifetimes and fluorescence spectra of fluorene oligomers of increasing length.

The code is suitable for investigating dynamical properties of large organic molecules, provided that a more reliable higher-level method
is used to filter those molecules out where tight-binding TD-DFT is too simplistic.

\section{Acknowledgements}
A.H. and R.M. acknowledge funding ERC Consolidator Grant DYNAMO (Grant No. 646737).

%%%%% APPENDICES %%%%%%%%%
\appendix

\section{Computational cost of long-range exchange and active space}
\label{sec:active_space}
Tight binding DFT has been designed for large systems that are out of reach with full DFT. For large system the charge transfer problem is particularly severe, so that some form of correction
 becomes mandatory. Unfortunately, the introduction of exact exchange partly destroys the efficiency of tight binding TD-DFT. The evaluation of the matrix product $(\mathbf{A} \pm \mathbf{B}) \vec{v}$ that comes up in the iterative solution of the TD-DFTB equations requires nested summations over orbital indeces $i,j,a,b$. Eqns. \ref{eqn:ApB} and \ref{eqn:ApBlr} show only the relevant parts of the summation with and without long-range exchange.
Without exact exchange the summations can be disentangled; the innermost sum $\sum_{jb} \left(q_B^{jb} v_{jb} \right)$ only depends on the atom index $B$:
\begin{equation}
\sum_A q_A^{ia} \left( \sum_B \gamma_{AB} \sum_{jb} \left( q_B^{jb} v_{jb} \right)\right) \quad\quad \label{eqn:ApB}
\end{equation}
The inclusion of exact exchange adds two additional terms, the first of them is
\begin{equation}
- \sum_A \sum_j q_A^{ij} \left( \sum_B \gamma_{AB}^{lr} \left( \sum_b q_B^{ab} v_{jb} \right)\right) \label{eqn:ApBlr}.
\end{equation}
The innermost sum $\sum_{j} q_B^{ja} v_{jb}$ still depends on three indices, $B$,$a$ and $b$. The computational effort becomes comparable to the full lc-TD-DFT equations with a minimal basis set.

One solution is to solve the TD-DFTB equations in a reduced active space: only single excitations from the highest $N_{\text{act. occ}}$ to the lowest $N_{\text{act. virt}}$ orbitals are considered. This approach is usually avoided in DFT calculations since many orbital transitions with low amplitude can still lower the energy considerably even if the excitation is dominated by a single orbital transition.
The excitation energies will be higher compared to the full active space, but the shape of the potential energy surfaces will be similar.

\begin{figure}[h!]
\includegraphics[width=0.8\textwidth]{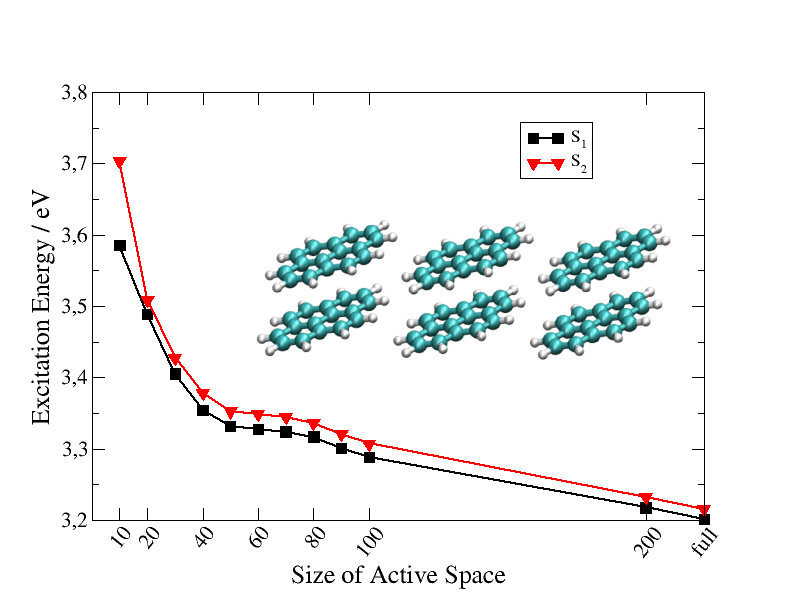}
\caption{Dependence of excitation energies on active space. The number of active occupied and virtual orbitals are marked on the horizontal axis.}
\label{fig:pyrene_3x2_active_space_inset}
\end{figure}

The effect of an active space on the excitation energies is visualized in Fig. \ref{fig:pyrene_3x2_active_space_inset}. 
From the crystal structure of pyrene\cite{pyrene_crystal_structure} 3 dimers with parallel molecular planes were selected. The molecular planes of the other nearest neighbor dimers are orthogonal so that the interaction is expected to be low. Surprisingly, the energy is still lowered by $0.1$ eV if the active space is increased from 100 active occupied and virtual orbitals to 200, although one would not expect excitations from HOMO$-100-x$ to LUMO$+100+x$ to be of any importance to the lowest excited state. This counter-intuitive effect should be kept in mind when restricting the space of excitations. In particular in non-adiabatic dynamics simulations it is tempting to use an active space as the speed-up allows to reach larger time scales. 

\section{Analytic gradients of ground and excited state energies}
\label{sec:analytic_gradients}

Efficient analytic gradients of TD-DFT excited states became first available with Furche's auxiliary functional method\cite{furche}, that avoids the time-consuming computation of gradients of the MO coefficients.
Chiba \cite{lc_gradients} adapted this idea to long-range corrected functionals. Heringer  
\cite{heringer} made the necessary simplifications needed to the tight-binding DFT and we now
complete this list with excited state gradients for long-range corrected tight-binding TD-DFT.

The following convention is used for orbital indices:
\begin{itemize}
\item p,q,r,s,t,u: general MO indices
\item i,j,k,l: occupied MO indices
\item a,b,c,d: virtual MO indices
\item Greek small letters: AO indices
\end{itemize}

An auxiliary functional\cite{furche,heringer} is defined that is variational in all arguments:
\begin{equation}
\begin{split}
L(X,Y,\Omega,C,Z,W) &= \frac{1}{2}\left\{(\vec{X}+\vec{Y}) (\boldsymbol{A}+\boldsymbol{B}) (\vec{X}+\vec{Y}) + (\vec{X}-\vec{Y}) (\boldsymbol{A}-\boldsymbol{B}) (\vec{X}-\vec{Y})\right\} - \Omega \left(\vec{X}^2 - \vec{Y}^2 - 1\right) \\
 & + \sum_{i,a} Z_{i a} H_{i a} - \sum_{p,q,p \leq q} W_{p q} (S_{p q} - \delta_{p q})
\end{split} \label{eqn:auxiliary_functional}
\end{equation}

\begin{eqnarray}
\frac{\partial L}{\partial \ket{X,Y}} &=& 0  \Rightarrow \text{ TD-DFT linear response equations } \\
\frac{\partial L}{\partial \Omega} &=& 0 \Rightarrow \text{ excitation vectors (X,Y) are orthonormal } \\
\frac{\partial L}{\partial \boldsymbol{Z}} &=& 0 \Rightarrow \text{ Kohn-Sham equations $H_{ia} = 0$ } \\
\frac{\partial L}{\partial \boldsymbol{W}} &=& 0 \Rightarrow \text{ Kohn-Sham orbitals are orthonormal }
\end{eqnarray}

The functional should also be stationary with respect to variations of the molecular orbital coefficients $C$, this requirement
determines the Lagrange multipliers $Z$ and $W$:

\begin{equation}
\frac{\partial L}{\partial \boldsymbol{C}} = 0 \Rightarrow \text{ determines $\boldsymbol{Z}$ and $\boldsymbol{W}$ } \label{eqn:dLdC}
\end{equation}

\subsection{Determination of the Lagrange multipliers}
%$i,j,k$ denote occupied orbitals, $a,b,c$ virtual orbitals and $p,q,r,s,t,u$ general orbitals.
Excited states with excitation energies $\Omega$ are the stationary points of the functional
\begin{equation}
G[X,Y,\Omega,C] = \frac{1}{2}\left\{(\vec{X}+\vec{Y}) (\boldsymbol{A}+\boldsymbol{B}) (\vec{X}+\vec{Y}) + (\vec{X}-\vec{Y}) (\boldsymbol{A}-\boldsymbol{B}) (\vec{X}-\vec{Y})\right\} - \Omega \left(\vec{X}^2 - \vec{Y}^2 -1 \right)
\end{equation}
which is part of the auxiliary functional $L$ in Eqn. \ref{eqn:auxiliary_functional}. The equations for the Lagrange multipliers are easier to deal with, if Eqn. \ref{eqn:dLdC} is transformed into
\begin{equation}
\left( \frac{\partial L}{\partial \boldsymbol{C}} \right)^T \boldsymbol{C} = 0 \quad \text{or componentwise} \quad \sum_{\mu} \frac{\partial L}{\partial C_{\mu p}} C_{\mu q} = 0.
\end{equation}

On the next few pages expressions for calculating
\begin{equation}
Q_{p q} = \sum_{\mu} \frac{\partial G}{\partial C_{\mu p}} C_{\mu q}
\end{equation}
are derived.

This envolves transforming the derivatives w/r/t the MO coefficients of the 0-th order Hamiltonian
\begin{equation}
\sum_{\mu} \frac{\partial H^0_{rs}}{\partial C_{\mu p}} C_{\mu q} = H^0_{q s} \delta_{p r} + H^0_{q r} \delta_{p s} 
\end{equation}
the overlap matrix
\begin{equation}
\sum_{\mu} \frac{\partial S_{rs}}{\partial C_{\mu p}} C_{\mu q} = S_{q s} \delta_{p r} + S_{q r} \delta_{p s} = \delta_{q s} \delta_{p r} + \delta_{q r} \delta_{p s}
\end{equation}
and the electron integrals
\begin{equation}
\sum_{\mu} \frac{\partial (r s | t u)}{\partial C_{\mu p}} C_{\mu q} = \delta_{p r} (q s | t u) + \delta_{p s} (r q | t u) + \delta_{p t} (r s | q u) + \delta_{p u} (r s | t q),
\end{equation}
for which the tight-binding approximations will be made later.

The Kohn-Sham Hamiltonian at the DFTB level with long-range correction reads:
\begin{equation}
H_{r s} = H^0_{r s} + \sum_{k \in occ} \left( 2 (r s | k k) - (r k | k s)_{\text{lr}} \right) \underbrace{- \sum_{\gamma,\delta} \left( (rs|\gamma\delta) - \frac{1}{2} (r\gamma|\delta s)_{\text{lr}} \right)P^0_{\gamma\delta}}_{\text{from reference density}}
\end{equation}

The transformed MO derivatives of the Hamiltonian are
\begin{equation}
\begin{split}
\sum_{\mu} \frac{\partial H_{r s}}{\partial C_{\mu p}} C_{\mu q} =& H^0_{q s} \delta_{p r} + H^0_{q r} \delta_{p s} \\
& + \sum_{k \in occ} 2 \left[\delta_{p r} (q s | k k) + \delta_{p s} (r q | k k) + \delta_{p k} (r s | k q) + \delta_{p k} (r s | q k) \right] \\
& - \sum_{k \in occ} \left[\delta_{p r} (q k | k s)_{\text{lr}} + \delta_{p s} (r k | k q)_{\text{lr}} + \delta_{p k} (r k | q s)_{\text{lr}} + \delta_{p k} (r q | k s)_{\text{lr}} \right] \\
& - \delta_{pr} \sum_{\gamma,\delta} \left( (qs|\gamma\delta) - \frac{1}{2} (q\gamma|\delta s) \right) P^0_{\gamma\delta} - \delta_{ps} \sum_{\gamma,\delta} \left( (qr|\gamma\delta) -\frac{1}{2} (q\gamma|\delta r)_{\text{lr}} \right) P^0_{\gamma\delta} \\
=& \quad \delta_{p r} \left\{ H^0_{q s} + \sum_{k \in occ} \left[2 (q s| k k) - (q k|k s)_{\text{lr}} \right] - \sum_{\gamma,\delta} \left( (qs|\gamma\delta) -\frac{1}{2} (q\gamma|\delta s)_{\text{lr}} \right) P^0_{\gamma\delta} \right\} \\
 &+\delta_{p s} \left\{ H^0_{q r} + \sum_{k \in occ} \left[2 (r q| k k) - (r k|k q)_{\text{lr}}\right] -\sum_{\gamma,\delta} \left( (qr|\gamma\delta) -\frac{1}{2} (q\gamma|\delta r)_{\text{lr}} \right) P^0_{\gamma\delta} \right\} \\
 &+ \delta(p \in occ) \left\{ 2 (r s | p q) - (r p| q s)_{\text{lr}} + 2 (r s | q p) - (r q | p s)_{\text{lr}} \right\} \\
=& \delta_{p r} H_{q s} + \delta_{p s} H_{q r} + \delta(p \in occ) \left\{ 4 (r s|p q) - (r p| q s)_{\text{lr}} - (r q | p s)_{\text{lr}} \right\} \\
=& \left(\delta_{p r} \delta_{q s} + \delta_{p s} \delta_{q r}  \right) \varepsilon_{q} + \delta(p \in \text{occ}) \left\{ 4 (r s | p q) - (r p | q s)_{\text{lr}} - (r q | p s)_{\text{lr}} \right\} \\
=& \left(\delta_{p r} \delta_{q s} + \delta_{p s} \delta_{q r}  \right) \varepsilon_{q} + \delta(p \in \text{occ}) \left( (A + B)_{rs,pq} - \delta_{p r} \delta_{q s} \left(\epsilon_{s} - \epsilon_{r}\right)\right) \\
%=& \left( \delta_{ps} \delta_{qr} + \delta_{pr} \delta_{qs}\right) \epsilon_r + \delta_{p \in \text{occ}} (A+B)_{rs,pq}
=& \delta_{p s} \delta_{q r} \epsilon_r + \delta_{pr} \delta_{qs} \left( \delta(p \in occ) \epsilon_p + \delta(p \in virt) \epsilon_q \right) + \delta(p \in occ) (A+B)_{rs,pq}.
\end{split}
\end{equation}

The $\mathbf{A}$ and $\mathbf{B}$ matrices for singlet states are
\begin{eqnarray}
^SA_{ia,jb} &=& \delta_{ij} H_{ab} - \delta_{ab} H_{ij} + 2 (ia|jb) - (ij|ab)_{\text{lr}} \\
^SB_{ia,jb} &=& 2 (ia|jb) - (ib|aj)_{\text{lr}}.
\end{eqnarray}
Adding and subtracting the A and B gives
\begin{eqnarray}
^S(A+B)_{ia,jb} &=& \delta_{ij} H_{ab} - \delta_{ab} H_{ij} + 4 (ia|jb) - (ij|ab)_{\text{lr}} - (ib|aj)_{\text{lr}} \\
^S(A-B)_{ia,jb} &=& \delta_{ij} H_{ab} - \delta_{ab} H_{ij} + (ib|aj)_{\text{lr}} - (ij|ab)_{\text{lr}},
\end{eqnarray}

The transformed MO derivatives of the sum and differences,
\begin{equation}
\begin{split}
\sum_{\mu} \frac{\partial (A+B)_{kc,ld}}{\partial C_{\mu p}} C_{\mu q} = &  \\
& \quad \delta_{kl} \epsilon_k \left(\delta_{pd} \delta_{qc} + \delta_{pc} \delta_{qd}\right) - \delta_{cd} \epsilon_c \left(\delta_{pk} \delta_{ql} + \delta_{pl} \delta_{qk}\right) \\
& + \delta_{pk} (A+B)_{qc,ld} + \delta_{pl} (A+B)_{kc,qd} + \delta_{pc} (A+B)_{kq,ld} + \delta_{pd} (A+B)_{kc,lq} \\
& + \delta_{cd} \delta_{pk} \delta_{ql} (\epsilon_l - \epsilon_k) + \delta_{p \in \text{occ}} \left(\delta_{kl} (A+B)_{cd,pq} - \delta_{cd} (A+B)_{kl,pq} \right),
\end{split}
\end{equation}
and 
\begin{equation}
\begin{split}
\sum_{\mu} \frac{\partial (A-B)_{kc,ld}}{\partial C_{\mu p}} C_{\mu q} = & \\
& \quad \delta_{kl} \epsilon_k \left(\delta_{pd} \delta_{qc} + \delta_{pc} \delta_{qd}\right) - \delta_{cd} \epsilon_c \left(\delta_{pl}\delta_{qk} + \delta_{pk}\delta_{ql}\right) \\
& + \delta_{pk} (A-B)_{qc,kd} + \delta_{pl} (A-B)_{kc,qd} + \delta_{pc} (A-B)_{kq,ld} + \delta_{pd} (A-B)_{kc,lq} \\
& + \delta_{cd} \delta_{pk} \delta_{ql} (\epsilon_l - \epsilon_k) + \delta_{p \in \text{occ}} \left(\delta_{kl} (A+B)_{cd,pq} - \delta_{cd} (A+B)_{kl,pq}\right),
\end{split}
\end{equation}

appear in the MO derivatives of the G functional

\begin{equation}
\begin{split}
Q_{pq} = \sum_{\mu} \frac{\partial G}{\partial C_{\mu p}} C_{\mu q} = \sum_{ia,jb} 
\frac{1}{2} \Big\{ 
\quad  &  (X+Y)_{ia} \left( \sum_{\mu} \frac{\partial (A+B)_{ia,jb}}{\partial C_{\mu p}} C_{\mu q} \right) (X+Y)_{jb} \\
+ & (X-Y)_{ia} \left( \sum_{\mu} \frac{\partial (A-B)_{ia,jb}}{\partial C_{\mu p}} C_{\mu q} \right) (X-Y)_{jb}
\Big\}
\end{split} \label{eqn:QpqXYABXY}
\end{equation}

To simplify Eqn. \ref{eqn:QpqXYABXY} the TD-DFT equations are exploited:

\begin{eqnarray}
\sum_{jb} (A+B)_{ia,jb} (X+Y)_{jb} &=& \Omega (X-Y)_{ia} \\
\sum_{ia} (X+Y)_{ia} (A+B)_{ia,jb} &=& \Omega (X-Y)_{jb} \\
\sum_{ib} (A-B)_{ia,jb} (X-Y)_{jb} &=& \Omega (X+Y)_{ia} \\
\sum_{ia} (X-Y)_{ia} (A-B)_{ia,jb} &=& \Omega (X+Y)_{jb}
\end{eqnarray}
Different cases have to be considered depending on whether the indeces $p,q$ belong to occupied or virtual orbitals:
\subsubsection{Case $p = i \in \text{occ}$, $q = j \in \text{occ}$}
\begin{equation}
\begin{split}
\sum_{\mu} \frac{\partial (A+B)_{kc,ld}}{\partial C_{\mu i}} C_{\mu j} = & \\
& - \delta_{cd} \epsilon_c \left(\delta_{ik} \delta_{jl} + \delta_{il} \delta_{jk}\right) \\
& + \delta_{ik} (A+B)_{jc,ld} + \delta_{il} (A+B)_{kc,jd} \\
& + \delta_{cd} \delta_{ik} \delta_{jl} (\epsilon_j - \epsilon_i) \\
& + \delta_{kl} (A+B)_{cd,ij} - \delta_{cd} (A+B)_{kl,ij}
\end{split}
\end{equation}

\begin{equation}
\begin{split}
\sum_{\mu} \frac{\partial (A-B)_{kc,ld}}{\partial C_{\mu i}} C_{\mu j} = & \\
& - \delta_{cd} \epsilon_c \left(\delta_{ik} \delta_{jl} + \delta_{il} \delta_{jk}\right) \\
& + \delta_{ik} (A-B)_{jc,kd} + \delta_{il} (A-B)_{kc,jd} \\
& + \delta_{cd} \delta_{ik} \delta_{jl} (\epsilon_j - \epsilon_i) \\
& + \delta_{kl} (A+B)_{cd,ij} - \delta_{cd} (A+B)_{kl,ij}
\end{split}
\end{equation}

Then
\begin{equation}
\begin{split}
Q_{ij} = 
& \quad \sum_{c} \Omega \left[(X+Y)_{ic} (X-Y)_{jc} + (X-Y)_{ic} (X+Y)_{jc}\right] \\
& - \sum_{c} \epsilon_c \left[(X+Y)_{ic} (X+Y)_{jc} + (X-Y)_{ic} (X-Y)_{jc} \right] \\
& + (\epsilon_j - \epsilon_i)\frac{1}{2}\sum_{c} \left[(X+Y)_{ic} (X+Y)_{jc} + (X-Y)_{ic} (X-Y)_{jc}\right] \\
& + \sum_{c,d} (A+B)_{ij,cd} \frac{1}{2} \sum_{k} \left[(X+Y)_{kc} (X+Y)_{kd} + (X-Y)_{kc} (X-Y)_{kd} \right] \\
& - \sum_{k,l} (A+B)_{ij,kl} \frac{1}{2} \sum_{c} \left[(X+Y)_{kc} (X+Y)_{lc} + (X-Y)_{kc} (X-Y)_{lc} \right]
\end{split}
\end{equation}

\subsubsection{Case $p = i \in \text{occ}$, $q = a \in \text{virt}$}
\begin{equation}
\begin{split}
\sum_{\mu} \frac{\partial (A+B)_{kc,ld}}{\partial C_{\mu i}} C_{\mu a} = & \\
& \quad \delta_{ik} (A+B)_{ac,ld} + \delta_{il} (A+B)_{kc,ad} \\
&   +   \delta_{kl} (A+B)_{cd,ia} - \delta_{cd} (A+B)_{kl,ia} 
\end{split}
\end{equation}

\begin{equation}
\begin{split}
\sum_{\mu} \frac{\partial (A-B)_{kc,ld}}{\partial C_{\mu i}} C_{\mu a} = & \\
& \quad \delta_{ik} (A-B)_{ac,ld} + \delta_{il} (A-B)_{kc,ad} \\
&   +   \delta_{kl} (A+B)_{cd,ia} - \delta_{cd} (A+B)_{kl,ia}
\end{split}
\end{equation}

Then
\begin{equation}
\begin{split}
Q_{ia} = 
&\quad \sum_{k,c,d} (A+B)_{ac,kd} (X+Y)_{ic} (X+Y)_{kd} \\
&  +   \sum_{k,c,d} (A-B)_{ac,kd} (X-Y)_{ic} (X-Y)_{kd} \\
&  +   \sum_{c,d} (A+B)_{ia,cd} \frac{1}{2} \sum_k \left[(X+Y)_{kc} (X+Y)_{kd} + (X-Y)_{kc} (X-Y)_{kd} \right] \\
&  -   \sum_{k,l} (A+B)_{ia,kl} \frac{1}{2} \sum_c \left[(X+Y)_{kc} (X+Y)_{lc} + (X-Y)_{kc} (X-Y)_{lc} \right]
\end{split}
\end{equation}

\subsubsection{Case $p = a \in \text{virt}$, $q = i \in \text{occ}$}
\begin{equation}
\sum_{\mu} \frac{\partial (A+B)_{kc,ld}}{\partial C_{\mu a}} C_{\mu i} = \delta_{ac} (A+B)_{ki,ld} + \delta_{ad} (A+B)_{kc,li} 
\end{equation}

\begin{equation}
\sum_{\mu} \frac{\partial (A-B)_{kc,ld}}{\partial C_{\mu a}} C_{\mu i} = \delta_{ac} (A-B)_{ki,ld} + \delta_{ad} (A-B)_{kc,li}
\end{equation}

Then
\begin{equation}
Q_{ai} = \sum_{k,l,c} (A+B)_{ki,lc} (X+Y)_{ka} (X+Y)_{lc} + \sum_{k,l,c} (A-B)_{ki,lc} (X-Y)_{ka} (X-Y)_{lc}
\end{equation}

\subsubsection{Case $p = a \in \text{virt}$, $q = b \in \text{virt}$}
\begin{equation}
\begin{split}
\sum_{\mu} \frac{\partial (A+B)_{kc,ld}}{\partial C_{\mu a}} C_{\mu b} = & 
\quad \delta_{kl} \epsilon_k \left(\delta_{ad} \delta_{bc} + \delta_{ac} \delta_{bd}\right) \\
&   +  \delta_{ac} (A+B)_{kb,ld} + \delta_{ad} (A+B)_{kc,lb}
\end{split}
\end{equation}

\begin{equation}
\begin{split}
\sum_{\mu} \frac{\partial (A-B)_{kc,ld}}{\partial C_{\mu a}} C_{\mu b} = &
\quad \delta_{kl} \epsilon_k \left(\delta_{ad} \delta_{bc} + \delta_{ac} \delta_{bd}\right) \\
&   +  \delta_{ac} (A-B)_{kb,ld} + \delta_{ad} (A-B)_{kc,lb}
\end{split}
\end{equation}

Then
\begin{equation}
\begin{split}
Q_{ab} = 
& \quad \sum_{k} \Omega \left[(X+Y)_{ka} (X-Y)_{kb} + (X-Y)_{ka} (X+Y)_{kb}\right]\\
& + \sum_{k} \epsilon_k \left[(X+Y)_{ka} (X+Y)_{kb} + (X-Y)_{ka} (X-Y)_{kb}\right]
\end{split}
\end{equation}

After defining the vectors
\begin{eqnarray}
U_{ab} &=& \sum_{i} \left[(X+Y)_{ia} (X-Y)_{ib} + (X-Y)_{ia} (X+Y)_{ib} \right] \\
U_{ij} &=& \sum_{a} \left[(X+Y)_{ia} (X-Y)_{ja} + (X-Y)_{ia} (X+Y)_{ja} \right] \\
V_{ab} &=& \sum_{i} \epsilon_i \left[(X+Y)_{ia} (X+Y)_{ib} + (X-Y)_{ia} (X-Y)_{ib} \right] \\
V_{ij} &=& \sum_{a} \epsilon_a \left[(X+Y)_{ia} (X+Y)_{ja} + (X-Y)_{ia} (X-Y)_{ja} \right] \\
T_{ab} &=& \frac{1}{2} \sum_i \left[(X+Y)_{ia} (X+Y)_{ib} + (X-Y)_{ia} (X-Y)_{ib} \right] \\
T_{ij} &=& \frac{1}{2} \sum_a \left[(X+Y)_{ia} (X+Y)_{ja} + (X-Y)_{ia} (X-Y)_{ja} \right]
\end{eqnarray}

one gets
\begin{align}
Q_{ij} &= \Omega U_{ij} - V_{ij} + (\epsilon_j - \epsilon_i) T_{ij} + \sum_{a,b} (A+B)_{ij,ab} T_{ab} - \sum_{k,l} (A+B)_{ij,kl} T_{kl} \\
\begin{split}
Q_{ia} &= \quad \sum_{c} (X+Y)_{ic} \sum_{k,d} (A+B)_{ac,kd} (X+Y)_{kd} + \sum_{c} (X-Y)_{ic} \sum_{k,d} (A-B)_{ac,kd} (X-Y)_{kd} \\
       & \quad   + \sum_{c,d} (A+B)_{ia,cd} T_{cd} - \sum_{k,l} (A+B)_{ia,kl} T_{kl}
\end{split} \\
Q_{ai} &= \sum_{k} (X+Y)_{ka} \sum_{l,c} (A+B)_{ki,lc} (X+Y)_{lc} + \sum_{k} (X-Y)_{ka} \sum_{l,c} (A-B)_{ki,lc} (X-Y)_{lc} \\
Q_{ab} &= \Omega U_{ab} + V_{ab}
\end{align}

Now, the DFTB approximations for two-electron integrals in terms of transition charges are used:
\begin{eqnarray}
(r s | t u) &=& \sum_{A,B} q_A^{rs} \gamma_{AB} q_B^{tu} \\
(r s | t u)_{\text{lr}} &=& \sum_{A,B} q_A^{rs} \gamma^{\text{lr}}_{AB} q_B^{tu} \\
\end{eqnarray}

We define the linear operators $H^+$ and $H^-$ (with the restriction on the indeces, $\delta_{pr} \delta_{qs} = 0$) and make use of the $\gamma$-approximation
for the electron integrals. The summation limits for $r, s$ depend on the nature of the vector $v_{rs}$.
\begin{equation}
\begin{split}
H^{+}_{pq}\left[v_{rs}\right] = & \sum_{r,s} (A+B)_{pq,rs} v_{rs} \\
=& \sum_{r,s} \left( 4 (pq|rs) - (pr|qs)_{\text{lr}} - (ps|qr)_{\text{lr}} \right) v_{rs} \\
=& \sum_{A,B} \sum_{r,s} \left( 4 q_A^{pq} \gamma_{AB} q_B^{rs} - q_A^{pr} \gamma_{AB}^{\text{lr}} q_B^{qs} - q_A^{ps} \gamma_{AB}^{\text{lr}} q_B^{qr} \right) v_{rs} \\
=& \quad 4 \sum_{A} q_A^{pq} \left(\sum_B \gamma_{AB} \left(\sum_{rs} \left(q_B^{rs} v_{rs}\right)\right)\right) \\
\cr & - \sum_{A} \sum_{r} q_A^{pr} \left(\sum_B \gamma_{AB}^{\text{lr}} \left(\sum_s q_B^{qs} v_{rs}\right)\right) 
- \sum_{A} \sum_{s} q_A^{ps} \left(\sum_B \gamma_{AB}^{\text{lr}} \left(\sum_r q_B^{qr} v_{rs}\right)\right)
\end{split}
\end{equation}
and
\begin{equation}
\begin{split}
H^{-}_{pq}\left[v_{rs}\right] = & \sum_{r,s} (A-B)_{pq,rs} v_{rs} \\
=& \sum_{r,s} \left( (ps|qr)_{\text{lr}} - (pr|qs)_{\text{lr}}\right) v_{rs} \\
=& \sum_{A,B} \sum_{r,s} \left( q_A^{ps} \gamma_{AB}^{\text{lr}} q_B^{qr} - q_A^{pr} \gamma_{AB}^{\text{lr}} q_B^{qs}\right) v_{rs} \\
=& \sum_{A} \sum_{s} q_A^{ps} \left(\sum_{B} \gamma_{AB}^{\text{lr}} \left(\sum_{r} q_B^{qr} v_{rs}\right)\right) 
  -\sum_{A} \sum_{r} q_A^{pr} \left(\sum_{B} \gamma_{AB}^{\text{lr}} \left(\sum_{s} q_B^{qs} v_{rs}\right)\right)
\end{split}
\end{equation}

and also 
\begin{equation}
\begin{split}
G_{ij} = &
(\epsilon_j - \epsilon_i) T_{ij} + \sum_{a,b} (A+B)_{ij,ab} T_{ab} - \sum_{k,l} (A+B)_{ij,kl} T_{kl} \\
=& \quad
    4 \sum_{A} q_A^{ij} \left(\sum_B \gamma_{AB} \left[\sum_{a,b} q_b^{ab} T_{ab} - \sum_{k,l} q_B^{kl} T_{kl}\right]\right) \\
&  +2 \sum_{A} \left(\sum_{k} q_A^{ik} \left(\sum_{B} \gamma_{AB}^{\text{lr}} \left(\sum_{l} q_B^{lj} T_{kl}\right)\right)\right) \\
&  -2 \sum_{A} \left(\sum_{a} q_A^{ia} \left(\sum_{B} \gamma_{AB}^{\text{lr}} \left(\sum_{b} q_B^{jb} T_{ab}\right)\right)\right) \\
=& H^{+}_{ij}[\vec{T}^{v-v}] - H^{+}_{ij}[\vec{T}^{o-o}]
\end{split}
\end{equation}

Finally one finds
\begin{eqnarray}
Q_{ij} &=& \Omega U_{ij} - V_{ij} + H^{+}_{ij}[\vec{T}^{v-v}] - H^{+}_{ij}[\vec{T}^{o-o}] \\ \label{eqn:Qij}
Q_{ia} &=& \quad \sum_{c} (X+Y)_{ic} H^{+}_{ac} \left[\vec{X}+\vec{Y}\right] + \sum_{c} (X-Y)_{ic} H^{-}_{ac} \left[\vec{X}-\vec{Y}\right] \\
 \cr &&       +  H^{+}_{ia}\left[\vec{T}^ {v-v}\right] - H^{+}_{ia}\left[\vec{T}^{o-o}\right] \\
Q_{ai} &=& \sum_{k} (X+Y)_{ka} H^{+}_{ki}\left[\vec{X}+\vec{Y}\right] + \sum_{k} (X-Y)_{ka} H^{-}_{ki}\left[\vec{X}-\vec{Y}\right] \\
Q_{ab} &=& \Omega U_{ab} + V_{ab} \label{eqn:Qab}
\end{eqnarray}

%--------------------------------------------------------------------------------------------------------
\rule{\linewidth}{1pt}

Now we need to find the equation for determining $Z$:
\begin{equation}
\sum_{\mu} \frac{\partial L}{\partial C_{\mu p}} C_{\mu q} = \underbrace{\sum_{\mu} \frac{\partial G}{\partial C_{\mu p}} C_{\mu q}}_{Q_{pq}} + \sum_{ia} Z_{ia} \sum_{\mu} \frac{\partial H_{ia}}{\partial C_{\mu p}} C_{\mu q} - \sum_{r,s,r\leq s} W_{rs} \sum_{\mu} \frac{\partial S_{rs}}{\partial C_{\mu p}} C_{\mu q} \stackrel{!}{=} 0
\end{equation}
The first term on the right hand side was determined above, Eqns. \ref{eqn:Qij}-\ref{eqn:Qab}, the second and third terms containing the sought for Lagrange multipliers Z and W are
\begin{equation}
\begin{split}
\sum_{ia} Z_{ia} \sum_{\mu} \frac{\partial H_{ia}}{\partial C_{\mu p}} C_{\mu q} &= \sum_{ia} Z_{ia} \left[\left(\delta_{pa} \delta_{qi} + \delta_{pi} \delta{qa}\right) \epsilon_i + \delta_{p \in \text{occ}} (A+B)_{ia,pq} \right] \\
&= Z_{qp} \epsilon_q + Z_{pq} \epsilon_p + \delta(p \in \text{occ}) \sum_{ia} Z_{ia} (A+B)_{ia,pq}
\end{split}
\end{equation}
and
\begin{equation}
\begin{split}
\sum_{r,s,r\leq s} W_{rs} \sum_{\mu} \frac{\partial S_{rs}}{\partial C_{\mu p}} C_{\mu q} &= \sum_{r,s,r\leq s} W_{rs} \left(\delta_{qs}\delta_{rp} + \delta_{qr}\delta_{sp}\right) \\
&= \sum_{r,s,r \leq s} \left(W_{pq} \delta_{qs} \delta_{pr} + W_{qp} \delta_{qr} \delta_{ps} \right) \\
&= 
\begin{cases}
W_{pq} & p < q \\
W_{qp} & p > q \\
W_{pq} + W_{qp} & p = q
\end{cases} \\
&= (1 + \delta_{pq}) W_{pq} \quad \quad \text{ since } W_{pq} = W_{qp}.
\end{split}
\end{equation}

This leads to the following equation for determining $Z$:
\begin{equation}
Q_{pq} + \left(Z_{qp} \epsilon_q + Z_{pq} \epsilon_p \right) + \delta_{p \in \text{occ}} \sum_{ia} Z_{ia} (A+B)_{ia,pq} = \left(1 + \delta_{pq}\right) W_{pq}
\end{equation}

The equation can be specialized for the occ-virt and the virt-occ blocks:
\begin{align}
Q_{ia} + Z_{ia} \epsilon_{i} + \sum_{jb} (A+B)_{ia,jb} Z_{jb} &= \left(1+\delta_{ia}\right) W_{ia} \quad \quad \text{ for } p \in \text{occ} = i \text{ and  } q \in \text{virt} = a \\
Q_{ai} + Z_{ia} \epsilon_{i} &= \left(1 + \delta_{ai} \right) W_{ai} \quad \quad \text{ for } p \in \text{virt} = a \text{ and } q \in \text{occ} = i
\end{align}
Subtracting the previous two equations gives (with $W_{ia} = W_{ai}$) the Z-vector equation:
\begin{equation}
\sum_{jb} (A+B)_{ia,jb} Z_{jb} = Q_{ai} - Q_{ia}
\end{equation}
The matrix becomes in the DFTB approximation:
\begin{equation}
(A+B)_{ia,jb} = \delta_{ij} \delta_{ab} (\epsilon_a - \epsilon_i) + 4 \sum_{A,B} q_A^{ia} \gamma_{AB} q_B^{jb} - \sum_{AB} q_A^{ij} \gamma_{AB}^{\text{lr}} q_B^{ab} - \sum_{A,B} q_A^{ib} \gamma_{AB}^{\text{lr}} q_B^{ja}
\end{equation}

\begin{equation}
\begin{split}
\sum_{k,b} (A+B)_{ij,kb} Z_{kb} = H^{+}_{ij}[\vec{Z}]
\end{split}
\end{equation}

After solving this system of linear equations for $Z$, the other Lagrange multiplier $W$ can be determined as
\begin{eqnarray}
W_{ij} &=& \frac{1}{1 + \delta_{ij}} \left( Q_{ij} + \sum_{k,b} (A+B)_{ij,kb} Z_{kb} \right) = \frac{1}{1 + \delta_{ij}} \left(Q_{ij} + H^{+}_{ij}[\vec{Z}] \right)\\
W_{ia} &=& W_{ai} = Q_{ai} + Z_{ia} \epsilon_i \\
W_{ab} &=& \frac{1}{1 + \delta_{ab}} Q_{ab}
\end{eqnarray}

\subsection{Assembling the gradient}
At the stationary point of $L$
\begin{equation}
L(X,Y,\Omega,C,Z,W) = \Omega \Rightarrow \frac{d L}{d R} = \frac{d \Omega}{d R}
\end{equation}
where $\frac{d }{d R}$ stands for the total derivative w/r/t an external parameter such as a nuclear coordinate.
Since $L$ is variational in all parameters,
\begin{equation}
\frac{d L}{d R} = 
 \cancelto{0}{\frac{\partial L}{\partial \ket{X,Y}}} \cdot \frac{\partial \ket{X,Y}}{\partial R} 
+\cancelto{0}{\frac{\partial L}{\partial \Omega}} \cdot \frac{\partial \Omega}{\partial R} 
+\cancelto{0}{\frac{\partial L}{\partial \boldsymbol{C}}} \cdot \frac{\partial \boldsymbol{C}}{\partial R}
+\cancelto{0}{\frac{\partial L}{\partial \boldsymbol{Z}}} \cdot \frac{\partial \boldsymbol{Z}}{\partial R}
+\cancelto{0}{\frac{\partial L}{\partial \boldsymbol{W}}} \cdot \frac{\partial \boldsymbol{W}}{\partial R}
+\frac{\partial L}{\partial R} 
\end{equation}
only the partial derivative $\frac{\partial L}{\partial R}$ survives.

The partial derivative of the G functional is
\begin{equation}
\begin{split}
\frac{\partial G}{\partial R} = 
 \frac{1}{2} \sum_{ia,jb} \Big[ & \quad
 (X+Y)_{ia} \left\{ \delta_{ij} \frac{\partial H_{ab}}{\partial R} - \delta_{ab} \frac{\partial H_{ij}}{\partial R} + 4 \frac{\partial (ia|jb)}{\partial R} - \frac{\partial (ij|ab)_{\text{lr}}}{\partial R} - \frac{\partial (ib|aj)_{\text{lr}}}{\partial R} \right\}  (X+Y)_{jb} \\
& + (X-Y)_{ia} \left\{ \delta_{ij} \frac{\partial H_{ab}}{\partial R} - \delta_{ab} \frac{\partial H_{ij}}{\partial R} + \frac{\partial (ib|aj)_{\text{lr}}}{\partial R} - \frac{\partial (ij|ab)_{\text{lr}}}{\partial R} \right\} (X-Y)_{jb} \Big] \\
= \frac{1}{2} \Big\{ 
 \sum_{a,b} & \frac{\partial H_{ab}}{\partial R} \sum_i \left[(X+Y)_{ia} (X+Y)_{ib} + (X-Y)_{ia} (X-Y)_{ib} \right] \\
-\sum_{i,j} & \frac{\partial H_{ij}}{\partial R} \sum_a \left[(X+Y)_{ia} (X+Y)_{ja} + (X-Y)_{ia} (X-Y)_{ja} \right] \\
+ 4 \sum_{ia,jb} & \frac{\partial (ia|jb)}{\partial R} (X+Y)_{ia} (X+Y)_{jb} \\
-   \sum_{ia,jb} & \frac{\partial (ij|ab)_{\text{lr}}}{\partial R} \left[(X+Y)_{ia} (X+Y)_{jb} + (X-Y)_{ia} (X-Y)_{jb} \right] \\
-   \sum_{ia,jb} & \frac{\partial (ib|aj)_{\text{lr}}}{\partial R} \left[(X+Y)_{ia} (X+Y)_{jb} - (X-Y)_{ia} (X-Y)_{jb} \right] \Big\}.
\end{split}
\end{equation}

The gradient of the excitation energy $\Omega$ is
\begin{equation}
\begin{split}
\frac{d \Omega}{d R} = & \frac{\partial G}{\partial R} + \sum_{ia} Z_{ia} \frac{\partial H_{ia}}{\partial R} - \sum_{p,q,p\leq q} W_{pq} \frac{\partial S_{pq}}{\partial R} \\
= & \quad \sum_{a,b} \frac{\partial H_{ab}}{\partial R} T_{ab} - \sum_{i,j} \frac{\partial H_{ij}}{\partial R} T_{ij} + \sum_{ia} \frac{\partial H_{ia}}{\partial R} Z_{ia} - \sum_{p,q,p \leq q} W_{pq} \frac{\partial S_{pq}}{\partial R} \\
  & + 2 \sum_{ia,jb} \frac{\partial (ia|jb)}{\partial R} (X+Y)_{ia} (X+Y)_{jb} \\
  & - \frac{1}{2} \sum_{ia,jb} \frac{\partial (ij|ab)_{\text{lr}}}{\partial R} \left[(X+Y)_{ia} (X+Y)_{jb} + (X-Y)_{ia} (X-Y)_{jb} \right] \\
  & - \frac{1}{2} \sum_{ia,jb} \frac{\partial (ib|aj)_{\text{lr}}}{\partial R} \left[(X+Y)_{ia} (X+Y)_{jb} - (X-Y)_{ia} (X-Y)_{jb} \right]. \label{eqn:dOmegadR}
\end{split}
\end{equation}
Now we switch to the AO basis, Greek letters $\alpha,\beta,\gamma,\delta$ denote atomic orbitals (AO).
Transforming the gradients w/r/t R of the overlap matrix into the AO basis gives
\begin{equation}
\frac{\partial S_{pq}}{\partial R} = \sum_{\alpha,\beta} C_{\alpha p} C_{\beta q} \frac{\partial S_{\alpha \beta}}{\partial R}.
\end{equation}

Remembering that the density matrix is defined as
\begin{equation}
P_{\gamma \delta} = 2 \sum_{k} C_{\gamma k} C_{\delta k},
\end{equation}
we find for the gradient of the Hamiltonian matrix in AO basis:
\begin{equation}
\begin{split}
\frac{\partial H_{pq}}{\partial R} =& \frac{\partial H^0_{pq}}{\partial R} + \sum_k \left(2 \frac{\partial (pq|kk)}{\partial R} - \frac{\partial (pk|kq)_{\text{lr}}}{\partial R} \right) - \sum_{\gamma\delta} \left( \frac{\partial (pq|\gamma\delta)}{\partial R} -\frac{1}{2} \frac{\partial (p\gamma|\delta q)_{\text{lr}}}{\partial R} \right) P^0_{\gamma\delta}\\
=& \sum_{\alpha,\beta} C_{\alpha p} C_{\beta q} \frac{\partial H^0_{\alpha\beta}}{\partial R} + \sum_{\alpha,\beta,\gamma,\delta} \frac{\partial (\alpha \beta | \gamma \delta)}{\partial R} C_{\alpha p} C_{\beta q} \left(\sum_k 2 C_{\gamma k} C_{\delta k}\right) \\
 & - \frac{1}{2} \sum_{\alpha,\beta,\gamma,\delta} \frac{\partial (\alpha \delta|\gamma \beta)_{\text{lr}}}{\partial R} C_{\alpha p} C_{\beta q} \left(\sum_k 2 C_{\gamma k} C_{\delta k} \right) \\
 & - \sum_{\alpha,\beta} C_{\alpha p} C_{\beta q} \left( \frac{\partial (\alpha\beta|\gamma\delta)}{\partial R} -\frac{1}{2} \frac{\partial (\alpha\delta|\gamma\beta)_{\text{lr}}}{\partial R} \right) P^0_{\gamma\delta} \\
=& \sum_{\alpha,\beta} C_{\alpha p} C_{\beta q} \underbrace{\left\{
\frac{\partial H^0_{\alpha\beta}}{\partial R} + \sum_{\gamma,\delta} \left(\frac{\partial (\alpha\beta|\gamma\delta)}{\partial R} \left(P_{\gamma\delta} - P^0_{\gamma\delta}\right) - \frac{1}{2} \frac{\partial (\alpha\delta|\gamma\beta)_{\text{lr}}}{\partial R}  \left(P_{\gamma\delta} - P^0_{\gamma\delta} \right)  \right)
\right\}}_{\frac{\partial H_{\alpha\beta}}{\partial R}}
\end{split}
\end{equation}
At this point we specify how the gradients of the electron integrals
look in the $\gamma$-approximation. $A$,$B$ enumerate atoms, $\alpha \in A$ means that the atomic orbital $\alpha$ is centered on atom $A$. The $\gamma$-matrix in the AO basis reads:
\begin{equation}
\gamma_{\alpha\beta} = \sum_{A,B} \gamma_{AB} \delta(\alpha \in A) \delta(\beta \in B)
\end{equation}

With the $\gamma$-approximation the Coulomb integrals in AO basis simplify to
\begin{equation}
(\alpha\beta|\gamma\delta) = \frac{1}{4} S_{\alpha\beta} S_{\gamma\delta} \left(\gamma_{\alpha\gamma} + \gamma_{\alpha\delta} + \gamma_{\beta\gamma} + \gamma_{\beta\delta} \right)
\end{equation}
and the electron integrals for long-range part of Coulomb potential simplify to
\begin{equation}
(\alpha\beta|\gamma\delta)_{\text{lr}} = \frac{1}{4} S_{\alpha\beta} S_{\gamma\delta} \left(\gamma^{\text{lr}}_{\alpha\gamma} + \gamma^{\text{lr}}_{\alpha\delta} + \gamma^{\text{lr}}_{\beta\gamma} + \gamma^{\text{lr}}_{\beta\delta} \right)
\end{equation}
with the gradients
\begin{equation}
\begin{split}
\frac{\partial (\alpha\beta|\gamma\delta)}{\partial R} = & \quad 
\frac{1}{4} \left(\frac{\partial S_{\alpha\beta}}{\partial R} S_{\gamma\delta} + S_{\alpha\beta} \frac{\partial S_{\gamma\delta}}{\partial R} \right) 
    \left[\gamma_{\alpha\gamma} + \gamma_{\alpha\delta} + \gamma_{\beta\gamma} + \gamma_{\beta\delta}\right] \\
& + 
\frac{1}{4} S_{\alpha\beta} S_{\gamma\delta} 
    \left[\frac{\partial \gamma_{\alpha\gamma}}{\partial R} + \frac{\partial \gamma_{\alpha\delta}}{\partial R} + \frac{\partial \gamma_{\beta\gamma}}{\partial R} + \frac{\partial \gamma_{\beta\delta}}{\partial R} \right]
\end{split} \label{eqn:electron_integrals_gradients}
\end{equation}
and a similar expression where $\gamma$ is replaced by $\gamma^{\text{lr}}$.

Next we will transform each term in Eqn. \ref{eqn:dOmegadR} separately into the AO basis:
\begin{itemize}
\item transform terms with two indeces
\begin{equation}
\sum_{a,b} \frac{\partial H_{ab}}{\partial R} T_{ab} = \sum_{\alpha\beta} \frac{\partial H_{\alpha\beta}}{\partial R} \underbrace{\sum_{a,b} C_{\alpha a} C_{\beta b} T_{ab}}_{T^{\text{v-v}}_{\alpha\beta}}
\end{equation}

\begin{equation}
\sum_{i,j} \frac{\partial H_{ij}}{\partial R} T_{ij} = \sum_{\alpha\beta} \frac{\partial H_{\alpha\beta}}{\partial R} \underbrace{\sum_{i,j} C_{\alpha i} C_{\beta j} T_{ij}}_{T^{\text{o-o}}_{\alpha\beta}}
\end{equation}

\begin{equation}
\sum_{i,a} \frac{\partial H_{ia}}{\partial R} Z_{ia} = \sum_{\alpha\beta} \frac{\partial H_{\alpha\beta}}{\partial R} \underbrace{\sum_{i,a} C_{\alpha i} C_{\beta a} Z_{ia}}_{Z_{\alpha\beta}}
\end{equation}

\begin{equation}
-\sum_{p,q,p\leq q} W_{pq} \frac{\partial S_{pq}}{\partial R} = - \sum_{\alpha\beta} \frac{\partial S_{\alpha\beta}}{\partial R} \underbrace{\sum_{p,q,p\leq q} C_{\alpha p} C_{\beta q} W_{pq}}_{W_{\alpha\beta}}
\end{equation}

\item transform Coulomb integrals which have 4 indeces
\begin{equation}
\begin{split}
&  2 \sum_{ia,jb} \frac{\partial (ia|jb)}{\partial R} (X+Y)_{ia} (X+Y)_{jb} = 2 \sum_{ia,jb} \sum_{\alpha,\beta,\gamma,\delta} \frac{\partial (\alpha\beta|\gamma\delta)}{\partial R} C_{\alpha i} C_{\beta a} C_{\gamma j} C_{\delta b} (X+Y)_{ia} (X+Y)_{jb} \\
=& 2 \sum_{\alpha,\beta,\gamma,\delta} \frac{\partial (\alpha\beta|\gamma\delta)}{\partial R} \underbrace{\left( \sum_{ia} C_{\alpha i} C_{\beta a} (X+Y)_{ia} \right)}_{(X+Y)_{\alpha\beta}} \left( \sum_{jb} C_{\gamma j} C_{\delta b} (X+Y)_{jb} \right) \\
=& 2 \sum_{\alpha,\beta,\gamma,\delta} \frac{\partial (\alpha\beta|\gamma\delta)}{\partial R} (X+Y)_{\alpha\beta} (X+Y)_{\gamma\delta}
\end{split}
\end{equation}

\item transform first long-range term
\begin{equation}
\begin{split}
 & -\frac{1}{2} \sum_{ia,jb} \frac{\partial (ij|ab)_{\text{lr}}}{\partial R} \left[(X+Y)_{ia} (X+Y)_{jb} + (X-Y)_{ia} (X-Y)_{jb}\right] \\
=& -\frac{1}{2} \sum_{\alpha,\beta,\gamma,\delta} \frac{\partial (\alpha\beta|\gamma\delta)_{\text{lr}}}{\partial R} 
\sum_{i,j,a,b} C_{\alpha i} C_{\beta j} C_{\gamma a} C_{\delta b} \left[(X+Y)_{ia} (X+Y)_{jb} + (X-Y)_{ia} (X-Y)_{jb} \right] \\
=& -\frac{1}{2} \sum_{\alpha,\beta,\gamma,\delta} \frac{\partial (\alpha\beta|\gamma\delta)_{\text{lr}}}{\partial R} \Big\{ \left(\sum_{ia} C_{\alpha i} C_{\gamma a} (X+Y)_{ia} \right) \left(\sum_{jb} C_{\beta j} C_{\delta b} (X+Y)_{jb} \right) \\
& \quad \quad \quad \quad \quad \quad \quad \quad \quad +\left(\sum_{ia} C_{\alpha i} C_{\gamma a} (X-Y)_{ia} \right) \left(\sum_{jb} C_{\beta j} C_{\delta b} (X-Y)_{jb}\right) \Big\} \\
=& -\frac{1}{2} \sum_{\alpha,\beta,\gamma,\delta} \frac{\partial (\alpha\beta|\gamma\delta)_{\text{lr}}}{\partial R} \left\{ (X+Y)_{\alpha\gamma} (X+Y)_{\beta\delta} + (X-Y)_{\alpha\gamma} (X-Y)_{\beta\delta} \right\}
\end{split}
\end{equation}
\item and similary the second long-range term
\begin{equation}
\begin{split}
& -\frac{1}{2} \sum_{ia,jb} \frac{\partial (ib|aj)_{\text{lr}}}{\partial R} \left[(X+Y)_{ia} (X+Y)_{jb} - (X-Y)_{ia} (X-Y)_{jb} \right] \\
=& -\frac{1}{2} \sum_{\alpha,\beta,\gamma,\delta} \frac{\partial (\alpha\beta|\gamma\delta)_{\text{lr}}}{\partial R} \left\{(X+Y)_{\alpha\gamma} (X+Y)_{\delta\beta} - (X-Y)_{\alpha\gamma} (X-Y)_{\delta\beta} \right\}.
\end{split}
\end{equation}

\end{itemize}

Everything put together, the gradient of the excitation energy becomes

\begin{equation}
\begin{split}
\frac{d \Omega}{d R} = & \quad 
\sum_{\alpha\beta} \frac{\partial H_{\alpha\beta}}{\partial R} 
  \left\{
T^{\text{v-v}}_{\alpha\beta} - T^{\text{o-o}}_{\alpha\beta} + Z_{\alpha\beta} 
  \right\} 
- \sum_{\alpha\beta} \frac{\partial S_{\alpha\beta}}{\partial R} W_{\alpha\beta} \\
& + 2 \sum_{\alpha,\beta,\gamma,\delta} \frac{\partial (\alpha\beta|\gamma\delta)}{\partial R} (X+Y)_{\alpha\beta} (X+Y)_{\gamma\delta} \\
& -\frac{1}{2} \sum_{\alpha,\beta,\gamma,\delta} \frac{\partial (\alpha\beta|\gamma\delta)_{\text{lr}}}{\partial R} 
 \Big\{
 (X+Y)_{\alpha\gamma} \left[(X+Y)_{\beta\delta} + (X+Y)_{\delta\beta}\right]
+(X-Y)_{\alpha\gamma} \left[(X-Y)_{\beta\delta} - (X-Y)_{\delta\beta}\right]
 \Big\}
\end{split}.
\end{equation}

Now we define two linear operators operating on a vector space with dimension
 $N_{\text{orb}} \times N_{\text{orb}}$: 

\begin{equation}
\begin{split}
\vec{F}_{\alpha\beta}[v] = & \sum_{\gamma,\delta} \frac{\partial (\alpha\beta|\gamma\delta)}{\partial R} v_{\gamma\delta} \\
= \frac{1}{4} \Big\{ & \quad 
\frac{\partial S_{\alpha\beta}}{\partial R} \left[\sum_{\gamma} \gamma_{\alpha\gamma} \left(\sum_{\delta} S_{\gamma\delta} \left(v_{\gamma\delta} + v_{\delta\gamma}\right)\right) + \sum_{\gamma} \left(\sum_{\delta} S_{\gamma\delta} \left(v_{\gamma\delta} + v_{\delta\gamma}\right)\right) \gamma_{\beta\gamma} \right]\\
& + S_{\alpha\beta} \Big[ \quad
\sum_{\gamma} \gamma_{\alpha\gamma} 
\left( \sum_{\delta} \frac{\partial S_{\gamma\delta}}{\partial R} \left(v_{\gamma\delta} + v_{\delta\gamma} \right) 
\right) 
+ \sum_{\gamma}\left( \sum_{\delta} \frac{\partial S_{\gamma\delta}}{\partial R} \left( v_{\gamma\delta} + v_{\delta\gamma} \right) 
\right)  \gamma_{\beta\gamma} \\
& \quad \quad \quad + \sum_{\gamma} \frac{\partial \gamma_{\alpha\gamma}}{\partial R} \left(\sum_{\delta} S_{\gamma\delta} \left(v_{\gamma\delta} + v_{\delta\gamma}\right)\right) + \sum_{\gamma}\left(\sum_{\delta} S_{\gamma\delta} \left(v_{\gamma\delta} + v_{\delta\gamma}\right)\right) \frac{\partial \gamma_{\beta\gamma}}{\partial R}\Big] \Big\}
\end{split}
\end{equation}

and 
\begin{equation}
\begin{split}
\vec{F}_{\alpha\beta}^{\text{lr}}[v] =  & \sum_{\gamma,\delta} \frac{\partial (\alpha\gamma|\beta\delta)_{\text{lr}}}{\partial R} v_{\delta\gamma} \\
 = \frac{1}{4} \Bigg\{ &
\quad  \gamma_{\alpha\beta}^{\text{lr}} \left(\sum_{\gamma} \frac{\partial S_{\alpha\gamma}}{\partial R} \left(\sum_{\delta} S_{\beta\delta} v_{\delta\gamma} \right)\right)
+ \sum_{\delta}\left(\left(\sum_{\gamma} \frac{\partial S_{\alpha\gamma}}{\partial R} v_{\delta\gamma}\right) \gamma_{\alpha\delta}^{\text{lr}}\right) S_{\beta\delta} \\
&+ \sum_{\gamma} \frac{\partial S_{\alpha\gamma}}{\partial R} \left(\left(\sum_{\delta} S_{\beta\delta} v_{\delta\gamma}\right) \gamma_{\beta\gamma}^{\text{lr}} \right)
+ \sum_{\gamma} \frac{\partial S_{\alpha\gamma}}{\partial R} \left(\sum_{\delta} S_{\beta\delta} \left(\gamma_{\delta\gamma}^{\text{lr}} v_{\delta\gamma}\right) \right) \\
&+ \gamma_{\alpha\beta}^{\text{lr}} \left(\sum_{\gamma} S_{\alpha\gamma} \left(\sum_{\delta} \frac{\partial S_{\beta\delta}}{\partial R} v_{\delta\gamma}\right)\right)
+ \sum_{\delta} \left(\left(\sum_{\gamma} S_{\alpha\gamma} v_{\delta\gamma}\right) \gamma_{\alpha\delta}^{\text{lr}}\right) \frac{\partial S_{\beta\delta}}{\partial R} \\
&+ \sum_{\gamma} S_{\alpha\gamma} \left(\left(\sum_{\delta} \frac{\partial S_{\beta\delta}}{\partial R} v_{\delta\gamma}\right) \gamma_{\beta\gamma}^{\text{lr}}\right)
+ \sum_{\gamma} S_{\alpha\gamma} \left(\sum_{\delta} \frac{\partial S_{\beta\delta}}{\partial R} \left(\gamma_{\delta\gamma}^{\text{lr}} v_{\delta\gamma}\right)\right) \\
&+ \frac{\partial \gamma_{\alpha\beta}^{\text{lr}}}{\partial R} \sum_{\gamma} S_{\alpha\gamma} \left(\sum_{\delta} S_{\beta\delta} v_{\delta\gamma}\right) 
+ \sum_{\delta} \left(\left(\sum_{\gamma} S_{\alpha\gamma} v_{\delta\gamma}\right) \frac{\partial \gamma_{\alpha\delta}^{\text{lr}}}{\partial R}\right) S_{\beta\delta} \\
&+ \sum_{\gamma} S_{\alpha\gamma} \left(\left(\sum_{\delta} S_{\beta\delta} v_{\delta\gamma}\right) \frac{\partial \gamma_{\beta\gamma}^{\text{lr}}}{\partial R}\right) 
+ \sum_{\gamma} S_{\alpha\gamma} \left(\sum_{\delta} \left(\frac{\partial \gamma_{\delta\gamma}^{\text{lr}}}{\partial R} v_{\delta\gamma}\right) S_{\beta\delta}\right) \Bigg\}
\end{split}
\end{equation}

Finally the gradient of the excitation energy becomes:
\begin{equation}
\begin{split}
\frac{d \Omega}{dR} = & \quad \sum_{\alpha,\beta} \frac{\partial H_{\alpha\beta}}{\partial R} \left\{T_{\alpha\beta}^{\text{v-v}} - T_{\alpha\beta}^{\text{o-o}} + Z_{\alpha\beta} \right\}
 - \sum_{\alpha,\beta} \frac{\partial S_{\alpha\beta}}{\partial R} W_{\alpha\beta} \\
& + 2 \sum_{\alpha,\beta} (X+Y)_{\alpha\beta} \vec{F}_{\alpha\beta}[(X+Y)_{\gamma\delta}] \\
& - \frac{1}{2} \sum_{\alpha,\beta} (X+Y)_{\alpha\beta} \vec{F}^{\text{lr}}_{\alpha\beta}[(X+Y)_{\gamma\delta} + (X+Y)_{\delta\gamma}] \\
& - \frac{1}{2} \sum_{\alpha,\beta} (X-Y)_{\alpha\beta} \vec{F}^{\text{lr}}_{\alpha\beta}[(X-Y)_{\delta\gamma} - (X-Y)_{\gamma\delta}]
\end{split}
\end{equation}

\subsection{Gradient of electronic energy of the ground state}

After defining the energy-weighted density matrix
\begin{equation}
P_{\alpha\beta}^{\text{en}} = 2 \sum_{k} \epsilon_k C_{\alpha k} C_{\beta k}
\end{equation}
the gradient of the ground state energy becomes:
\begin{equation}
\frac{d E_0}{dR} = \sum_{\alpha\beta} \left( \frac{\partial H_{\alpha\beta}^0}{\partial R} P_{\alpha\beta} 
+ \frac{1}{2} \left( \vec{F}_{\alpha\beta}[P-P_0] \left(P_{\alpha\beta} - P^0_{\alpha\beta}\right) - \frac{1}{2} \vec{F}^{\text{lr}}_{\alpha\beta}[P-P_0] \left(P_{\alpha\beta} - P^0_{\alpha\beta}\right)\right)
- \frac{\partial S_{\alpha\beta}}{\partial R} P_{\alpha\beta}^{\text{en}} \right)
\end{equation}
In the Coulomb part the density difference, $P-P_0$, has to be used because the gradient belonging to the reference density $P_0$ is already contained in $\frac{\partial H^0_{\alpha\beta}}{\partial R}$.

%%%%%%%%%%%%%%%%%%%%%%
\subsection{gamma-matrices}
Here we give expressions for the $\gamma$-matrices that are required for calculating electron integrals and their gradients in Eqn. \ref{eqn:electron_integrals_gradients}.
For charge fluctuations that have the form of Gaussians, the $\gamma$-matrix becomes:
\begin{equation}
\gamma_{AB} = \frac{\erf\left(C_{AB} R\right)}{R}.
\end{equation}
$R$ is the distance between the atomic centers $A$ and $B$ and 
\begin{equation}
C_{AB} = \frac{1}{\sqrt{2 \left(\sigma_A^2 + \sigma_B^2\right)}}
\end{equation}
depends on the widths $\sigma_A$ and $\sigma_B$ of the charge clouds on the two atoms. The widths are determined by the atom-specific Hubbard parameters $U_A$ as
\begin{equation}
\sigma_A = \frac{1}{\sqrt{\pi} U_A}.
\end{equation}
The long-range $\gamma$-matrix has the same form,
\begin{equation}
\gamma_{AB}^{\text{lr}} = \frac{\erf\left(C^{\text{lr}}_{AB} R\right)}{R},
\end{equation}
where 
\begin{equation}
C_{AB}^{\text{lr}} = \frac{1}{\sqrt{2 \left( \sigma_A^2 + \sigma_B^2 + \frac{1}{2} R_{\text{lr}}^2 \right)}}
\end{equation}
depends on the range-separation parameter $R_{\text{lr}}$. 

%%%%%%%%%%%%%%%%%%%%%%%

\section{Scalar non-adiabatic couplings}
\label{sec:scalar_couplings}
$\phi_i$ denotes the $i$-th spatial KS orbital at time $t$ and $\chi_j$
the $j$-th spatial KS orbital at time $t+ \Delta t$. $\alpha$ and $\beta$
denote the spin functions. $\ket{\{ \text{orbitals} \}}$ represents a Slater determinant of spin or spatial orbitals. $i,j$ label occupied orbitals, $a,b$ virtual orbitals and $r,s$ general orbitals.
In the DFTB ground state of a closed shell molecule with $N$ electrons the lowest $N/2$ spatial orbitals are doubly occupied:
\begin{equation}
\ket{\Psi_0^S(t)} = \ket{\{\phi_1\alpha, \ldots, \phi_i\alpha,\ldots,\phi_{N/2}\alpha; \phi_1\beta,\ldots,\phi_i\beta,\ldots,\phi_{N/2}\beta\}}
\end{equation}

The singly excited spin paired configuration state functions are generated by applying the
excitation operator
\begin{equation}
\hat{E}^S_{ia} = \frac{1}{\sqrt{2}} \left(\hat{a}^{\dagger}_{a\alpha} \hat{a}_{i\alpha} + \hat{a}^{\dagger}_{a\beta} \hat{a}_{i\beta} \right)
\end{equation}
to the ground state wavefunction:
\begin{equation}
\begin{split}
\ket{\Psi^S_{ia}(t)} = \hat{E}^S_{ia} \ket{\Psi_0(t)} = \frac{1}{\sqrt{2}} \Big[
 & \quad \ket{\{\phi_1\alpha,\ldots,\phi_{i-1}\alpha,{\color{red} \phi_{a}\alpha},\phi_{i+1}\alpha,\ldots,\phi_{N/2}\alpha; \phi_1\beta,\ldots,\phi_i\beta,\ldots,\phi_{N/2}\beta \} } \\
 &    +  \ket{\{\phi_1\alpha,\ldots,\phi_i\alpha,\ldots,\phi_{N/2}\alpha; \phi_1\beta,\ldots,\phi_{i-1}\beta,{\color{red} \phi_a\beta},\phi_{i+1}\beta,\ldots,\phi_{N/2}\beta \}} 
 \Big]
\end{split}
\end{equation}

The overlap between two singly excited configuration state functions at different times becomes
\begin{equation}
\begin{split}
\bracket{\Psi_{ia}^S(t)}{\Psi_{jb}^S(t+\Delta t)} = \frac{1}{2}  \Big( & \quad \bra{\{\phi_1\alpha,\ldots,{\color{red} \phi_a\alpha},\ldots,\phi_{N/2}\alpha; \phi_1\beta,\ldots,\phi_i\beta,\ldots,\phi_{N/2}\beta \}} \\
       & + \bra{\{\phi_1\alpha,\ldots,\phi_i\alpha,\ldots,\phi_{N/2}\alpha; \phi_1\beta,\ldots,{\color{red} \phi_a\beta},\ldots,\phi_{N/2}\beta \}} \Big) \\
 \times \Big( & \quad \ket{\{\chi_1\alpha,\ldots,{\color{red} \chi_b\alpha},\ldots,\chi_{N/2}\alpha; \chi_1\beta,\ldots,\chi_j\beta,\ldots,\chi_{N/2}\beta \}} \\
              & + \ket{\{ \chi_1\alpha,\ldots,\chi_j\alpha,\ldots,\chi_{N/2}\alpha; \chi_1\beta,\ldots,{\color{red} \chi_b\beta},\ldots,\chi_{N/2}\beta \}} \Big).
\end{split}
\end{equation}
Since $\bracket{\alpha}{\alpha} = 1$ and $\bracket{\alpha}{\beta} = 0$, the resulting determinants contain two blocks on the diagonal, one for $\alpha-\alpha$, the other for $\beta-\beta$. The same products appear twice, therefore one gets:

\begin{equation}
\resizebox{0.9\hsize}{!}{$
\begin{split}
\bracket{\Psi_{ia}^S(t)}{\Psi_{jb}^S(t+\Delta t)} = &
 \quad \bracket{\phi_1,\ldots,{\color{red} \phi_a},\ldots,\phi_{N/2}}{\chi_1,\ldots,{\color{red} \chi_b},\ldots,\chi_{N/2}} \bracket{\phi_1,\ldots,\phi_i,\ldots,\phi_{N/2}}{\chi_1,\ldots,\chi_j,\ldots,\chi_{N/2}} \\
&    +  \bracket{\phi_1,\ldots,{\color{red} \phi_a},\ldots,\phi_{N/2}}{\chi_1,\ldots,\chi_j,\ldots,\chi_{N/2}} \bracket{\phi_1,\ldots,\phi_i,\ldots,\phi_{N/2}}{\chi_1,\ldots,{\color{red} \chi_b},\ldots,\chi_{N/2}}
\end{split}
$}
\end{equation}
The overlap between two Slater determinants built from different sets of orbitals can be calculated as the determinant 
\begin{equation}
\bracket{\phi_1,\ldots,\phi_r,\ldots}{\chi_1,\ldots,\chi_s,\ldots} = 
\det
\begin{pmatrix}
\bracket{\phi_1}{\chi_1} & \hdots & \bracket{\phi_1}{\chi_s} & \hdots \\
\vdots                   & \hdots & \vdots                   & \hdots \\
\bracket{\phi_r}{\chi_1} & \hdots & \bracket{\phi_r}{\chi_s} & \hdots \\
\vdots                   & \hdots & \vdots                   & \hdots 
\end{pmatrix} \label{eqn:slater_overlap}.
\end{equation}

Denoting the overlap matrix between KS orbitals at different geometries by
\begin{equation}
S^{\text{mo}}_{r,s}(t,t+\Delta t) = \bracket{\phi_r}{\chi_s}
\end{equation}
the overlap in Eqn. \ref{eqn:slater_overlap} can be obtained as the determinant of a $(N/2 \times N/2)$-dimensional submatrix of $S^{\text{mo}}$:
\begin{equation}
 \det\left( S_{(\text{rows } 1,\ldots,r,\ldots),(\text{columns } 1,\ldots,s,\ldots)} \right)
\end{equation}

%%%% REFERENCES %%%%

%\bibliographystyle{achemso}
%\bibliographystyle{pccp}
\bibliographystyle{ieeetr}
\bibliography{references}

\end{document}